\documentclass[altaffilletter,aps,nofootinbib,twocolumn,prd,eqsecnum,preprintnumbers,superscriptaddress]{revtex4-1}
\pdfoutput=1
\usepackage[caption=false]{subfig}
\usepackage{graphicx}
\usepackage{amsmath}
\usepackage{amsfonts}
\usepackage{mathtools}
\usepackage{amssymb}
\usepackage{color}
\usepackage{bm}
\usepackage{mathrsfs}
\usepackage{epstopdf}
\usepackage{url}
\usepackage{footnote}
\usepackage{textcomp}
\usepackage{dsfont}
\usepackage{ulem}
\usepackage{hyperref}
\usepackage{enumerate}

\makeatletter
\newcommand*{\rom}[1]{\expandafter\@slowromancap\romannumeral #1@}
\makeatother

\begin{document}

\title{A consistent model of non-singular Schwarzschild black hole in loop quantum gravity and its quasinormal modes}

\author{Mariam Bouhmadi-L\'{o}pez}
\email{mariam.bouhmadi@ehu.eus}
\affiliation{Department of Theoretical Physics, University of the Basque Country UPV/EHU, P.O. Box 644, 48080 Bilbao, Spain}
\affiliation{IKERBASQUE, Basque Foundation for Science, 48011, Bilbao, Spain}

\author{Suddhasattwa Brahma}
\email{suddhasattwa.brahma@gmail.com}
\affiliation{Asia Pacific Center for Theoretical Physics, Pohang 37673, Republic of Korea}
\affiliation{Department of Physics, McGill University, Montr\'eal, QC H3A 2T8, Canada}

\author{Che-Yu Chen}
\email{b97202056@gmail.com}
\affiliation{Department of Physics and Center for Theoretical Sciences, National Taiwan University, Taipei, Taiwan 10617}
\affiliation{LeCosPA, National Taiwan University, Taipei, Taiwan 10617}

\author{Pisin Chen}
\email{pisinchen@phys.ntu.edu.tw}
\affiliation{Department of Physics and Center for Theoretical Sciences, National Taiwan University, Taipei, Taiwan 10617}
\affiliation{LeCosPA, National Taiwan University, Taipei, Taiwan 10617}
\affiliation{Kavli Institute for Particle Astrophysics and Cosmology, SLAC National Accelerator Laboratory, Stanford University, Stanford, CA 94305, USA}

\author{Dong-han Yeom}
\email{innocent.yeom@gmail.com}
\affiliation{Department of Physics Education, Pusan National University, Busan 46241, Republic of Korea}
\affiliation{Research Center for Dielectric and Advanced Matter Physics, Pusan National University,
Busan 46241, Republic of Korea}

\begin{abstract}
\noindent We investigate the interior structure, perturbations, and the associated quasinormal modes of a quantum black hole model recently proposed by Bodendorfer, Mele, and M\"unch (BMM). Within the framework of loop quantum gravity, the quantum parameters in the BMM model are introduced through polymerization, consequently replacing the Schwarzschild singularity with a spacelike transition surface. By treating the quantum geometry corrections as an `effective' matter contribution, we first prove the violation of energy conditions (in particular the null energy condition) near the transition surface and then investigate the required junction conditions on it. In addition, we study the quasinormal modes of massless scalar field perturbations, electromagnetic perturbations, and axial gravitational perturbations in this effective model. As expected, the quasinormal spectra deviate from their classical counterparts in the presence of quantum corrections. Interestingly, we find that the quasinormal frequencies of perturbations with different spins share the same qualitative tendency with respect to the change of the quantum parameters in this model.  
\end{abstract}

\maketitle

\section{Introduction}   
 
A general consensus is that general relativity (GR) has been tremendously successful in describing our universe. Based on GR, we have the big bang theory that stands as the building block of current standard cosmology. More recently, direct detections of gravitational waves from binary merger events \cite{Abbott:2016blz,Abbott:2017oio,TheLIGOScientific:2017qsa} and the image of the black hole shadow \cite{Akiyama:2019cqa} provide further evidences regarding the correctness of GR in the large curvature regime, such as that in the vicinity of neutron stars and black holes. Especially, the direct detection of gravitational waves has ushered in a new era of modern astronomy because we are now equipped with a \textit{telescope} more powerful than ever before to look much deeper into our universe, or to probe what really goes on very close to a black hole. 
 
However, GR, in spite of its remarkable successes, is incomplete from a theoretical point of view. For example, GR predicts the existence of spacetime singularities where the theory itself ceases to be valid \cite{Penrose:1964wq,Hawking:1969sw}. According to GR, there exists a singularity inside the black hole, while the hope is that the singularity can be ameliorated by some quantum effects. To address issues such as this, one major challenge is to figure out how a consistent quantum theory of gravity can be formulated. In recent decades, several approaches towards quantum gravity have been developed, such as loop quantum gravity (LQG) \cite{Rovelli:2011eq,Bojowald:2008zzb,Ashtekar:2013hs}, canonical quantum gravity \cite{qgkiefer}, string theory \cite{Strominger:1996sh}, and Euclidean path integral \cite{Hartle:1983ai,EQG,Chen:2018aij}. Since none of these approaches are complete by themselves, a more phenomenological perspective suggests thinking about the kind of effective quantum corrections that may contribute to the modifications to GR when considering regimes with large curvature scales. Such modified theories of gravity could be treated as effective, or semiclassical, approximations of the unknown full quantum theory of gravity \cite{Capozziello:2011et}.

Among the plethora of various approaches, in this paper we will consider the LQG scenario and focus on an effective black hole model within this approach. LQG introduces the concept of quantum geometry in the sense that the spacetime is made of some fundamental building blocks (known as spin-networks) which have a discrete spectrum for geometrical operators such as volume and area, with a so-called minimum area-gap \cite{Rovelli:2011eq}. The general idea behind constructing effective models in LQG is that one expects some non-perturbative quantum geometry corrections that would modify Einstein equations. One expects to write down such semiclassical  corrections which arise when considering the full quantum theory with expectation values of operators taken in some well-defined semiclassical states \cite{Bojowald:2012sp}. In practice, for effective versions of LQG, one typically introduces the so-called holonomy modifications by hand, which are quantum corrections resulting from the above-mentioned area-gap and the structure of the Hilbert space in LQG \cite{Ashtekar:1995zh,Ashtekar:1996eg,Ashtekar:2003hd}. 

The typical method of introducing holonomy modifications in effective models is to use the polymerization technique in LQG, which is achieved by replacing the conjugate momenta $p$ in the phase space with their polymerized counterparts $\sin(\lambda p)/\lambda$, where $\lambda$ is a quantum parameter (related to the area-gap) in the effective model. The trigonometric function is not an \textit{ad hoc} choice, but appears as the result of having matrix elements of SU($2$) holonomies evaluated along a loop. The curvature of spacetime, calculated in terms of these holonomies, gets a natural regularization as a result of using such bounded functions. When applied to cosmology, it is well-known that the big bang singularity is resolved\footnote{Whether one gets a primordial bounce \cite{Ashtekar:2011ni} or a signature-change \cite{Bojowald:2015gra} remains a matter of considerable debate.} \cite{Bojowald:2008zzb}. Using similar methods, one can also consider effective models to describe black hole spacetimes in LQG \cite{BenAchour:2016brs,Brahma:2014gca,Ashtekar:2005qt,Modesto:2005zm,Campiglia:2007pr,Modesto:2008im,Gambini:2013ooa,Bojowald:2016itl,Bohmer:2007wi,Chiou:2008nm,Chiou:2008eg,Joe:2014tca,Hossenfelder:2009fc,BenAchour:2018khr,Ashtekar:2018lag,Ashtekar:2018cay,Bojowald:2018xxu,Alesci:2018loi}. In spite of the various possibilities to construct effective black hole models, a common feature is the presence of a transition surface inside the black hole, which replaces the classical singularity, although many of them may suffer from other fundamental inconsistencies \cite{Brahma:2018cgr,Bodendorfer:2019xbp,Bouhmadi-Lopez:2019hpp,Bojowald:2019dry,BenAchour:2020gon,BenAchour:2020bdt,BenAchour:2020mgu}. Finally, when $\lambda p\ll1$, the classical theory can be easily recovered.

In this paper, we will focus on the effective quantum black hole proposed in Ref.~\cite{Bodendorfer:2019cyv}, which we shall refer to as the BMM model. Instead of the SU($2$) connections and their conjugate momenta, which are commonly used in LQG, the BMM model is based on the polymerization of a new set of canonical phase space variables. This model is characterized by the following features:
\begin{enumerate}[(i)]
\item The singularity is replaced with a spacelike transition surface, which separates infinite pairs of trapped and anti-trapped regions.
\item Quantum effects become important starting from a unique curvature scale which is independent of the mass of the black hole.
\item The solution recovers the Schwarzschild black hole near the event horizon and it is asymptotically flat.
\item A certain mass (de-)amplification relation is required.
\end{enumerate}
We will first investigate in more details the interior structure of the BMM quantum black hole, including the violation of energy conditions near the transition surface, and then examine the junction conditions on the transition surface. By treating the quantum corrections on the spacetime as a kind of effective matter fields, we confirm the expectation that the energy conditions (in particular the null energy condition) must be violated as a result of the \textit{repulsive} behavior near the transition surface, which prevents the formation of singularities. In addition, we will adopt the Israel junction conditions for a spacelike hypersurface embedded in the most general static and spherically symmetric spacetime{\footnote{Note that the existence of a spacelike transition surface is a generic property, i.e., not only in the BMM black hole model, but also for most black holes bouncing into a white hole in LQG. The junction conditions can be derived in the framework of GR. However, this does not invalidate their use in this model since in an effective approach to LQG, it is customary to rewrite the quantum geometry corrections as an ``effective" energy-momentum tensor coupled to GR. We will discuss this issue in more detail later.}}. It should be mentioned that the junction condition and the equations of motion for general shells of arbitrary causal character (which may even change) have been deduced and studied in arbitrary dimension \cite{Mars:1993mj,Mars:2007wy,Senovilla:2018hrw,Mars:2000gu,Mars:2007zt}, where the Israel junction conditions are generalized. We will then study the junction conditions on the transition surface of the BMM model and show that the transition between the black hole and the white hole regimes is perfectly smooth.

Next, we will study the perturbations of the BMM quantum black hole and calculate their ringing frequencies. In light of the recent detection of gravitational waves from binary merger events, the black hole perturbations and their associated gravitational wave properties have been intensively investigated \cite{Nollert:1999ji, Berti:2009kk, Konoplya:2011qq, Berti:2015itd, Cardoso:2019mqo, McManus:2019ulj}. Essentially, during the post-merger epoch and before the settlement of the new black hole, there is an intermediate stage where the distortion of the black hole can be revealed. This process is accompanied with the emission of gravitational waves and is dubbed as the ringdown stage. The ringdown signals are featured by quasinormal modes (QNMs) since the black hole at this stage can be treated as a dissipative system, continuously losing its energy. The perturbations have a discrete spectrum and the QNMs have complex frequencies, which are characterized only by the black hole mass, charge, and spin. If there are additional parameters that describe the black hole, such as some quantum parameters, they shall also leave their signatures in the QNM spectra.   

Typically, the QNMs of a black hole are described by a family of master equations, which are derived either by perturbing the gravitational equation directly or by considering the conservation equations of some test fields around the black hole spacetime. The former corresponds to the gravitational perturbations of the black hole and they can be further categorized into axial and polar perturbations, according to how the perturbations react to the change of parity. The latter, on the other hand, describes the evolution of the test fields around the black hole. For example, one can study the QNMs of a test scalar field and electromagnetic fields by starting with the Klein-Gordon equation and the Maxwell equations, respectively. Even though the latter case seems less related to the real gravitational wave signals, it nevertheless may provide us the first insight encoded in QNMs, such as the stability of the black hole.

In this work, we will study the QNMs of the massless scalar field perturbations, electromagnetic perturbations, and the axial gravitational perturbations in the BMM quantum black hole. We will review the derivation of the master equations of the massless scalar field and the electromagnetic fields, for the most general static and spherically symmetric spacetime. As for the axial gravitational perturbations, we will assume that the perturbed BMM black hole can be described within GR framework with an effective energy-momentum tensor endowed with an anisotropic fluid. The master equation is then obtained by perturbing Einstein equation and the effective energy-momentum tensor. For each case, the QNM frequencies are calculated with the Wentzel-Kramers-Brillouin (WKB) method up to the 6th order \cite{Schutz:1985zz,Iyer:1986np,Konoplya:2003ii,Matyjasek:2017psv,Konoplya:2019hlu,Matyjasek:2019eeu} as well as the asymptotic iteration method (AIM) \cite{Cho:2009cj,Cho:2011sf}. We will compare the results with those of the Schwarzschild black hole and demonstrate how the QNM frequencies change with the quantum parameters. In the context of LQG, the QNMs of the scalar field perturbations \cite{Chen:2011zzi,Santos:2015gja} and the axial gravitational perturbations \cite{Cruz:2015bcj} for the self-dual black hole \cite{Modesto:2008im} have been studied. The shadow and QNMs of rotating self-dual black holes have also been investigated very recently \cite{Liu:2020ola}.

This paper is outlined as follows. In section~\ref{sec.bmmbh}, we review the BMM quantum black hole, including its derivation and some of its important features. This section gives a summary of the BMM quantum black hole, and the content of our new results is included thereafter. In section~\ref{sec.ts}, we investigate physical properties near the interior transition surface, such as the violation of energy conditions and the implication of the junction conditions. In section~\ref{sec.QNMbig}, we study the massless scalar field perturbations, electromagnetic perturbations, and the axial gravitational perturbations of the BMM quantum black hole. The QNM frequencies are evaluated and they are compared with those of the Schwarzschild black hole in GR. We finally conclude in section \ref{conclu}. An appendix \ref{WKBapp} is included to give a brief introduction on the WKB method applied in this work.

\section{BMM quantum black hole}\label{sec.bmmbh}
In this section, we briefly review the effective LQG black hole (the BMM black hole) proposed in Ref.~\cite{Bodendorfer:2019cyv}, including its derivation and its important features. This effective model is based on the polymerization of a new set of canonical phase space variables, rather than the standard connection variables that have been used in the majority of LQG effective black hole models. We will briefly review the classical spacetime delineated by the Hamiltonian description and then review the effective black hole model proposed in Ref.~\cite{Bodendorfer:2019cyv}.

\subsection{The Hamiltonian description: classical theory}
In order to consider the black hole spacetime, we start with the static and spherically symmetric metric 
\begin{equation}
ds^2=-\bar{a}(r)dt^2+N(r)dr^2+\bar{b}(r)^2d\Omega_2^2\,,\label{SSSeq1}
\end{equation}
where $\bar{a}(r)$, $N(r)$, and $\bar{b}(r)$ are functions of $r$. Inserting the line element \eqref{SSSeq1} into the Einstein-Hilbert action, one can derive the effective Lagrangian as follows
\begin{equation}
L(\bar{a},\bar{b},\bar{n})=2L_0\sqrt{\bar{n}}\left(\frac{\bar{a}'\bar{b}'\bar{b}}{\bar{n}}+\frac{\bar{a}\bar{b}'^2}{\bar{n}}+1\right)\,,
\end{equation} 
where the prime denotes the derivative with respect to $r$ and the function $\bar{n}$ is given by
\begin{equation}
\bar{n}=\bar{a}N(r)\,.\nonumber
\end{equation}
The constant $L_0$ is an infrared cut-off in the non-compact $t$ direction ($0\le t\le L_0$) for a fiducial cell in the constant $r$ slices \cite{Bodendorfer:2019cyv}. After redefining the variables 
\begin{equation}
\sqrt{n}=L_0\sqrt{\bar{n}}\,,\qquad\sqrt{a}=L_0\sqrt{\bar{a}}\,,\qquad b=\bar{b}\,,
\end{equation}
the Lagrangian can be rewritten as
\begin{equation}
L(a,b,n)=2\sqrt{n}\left(\frac{a'b'b}{n}+\frac{ab'^2}{n}+1\right)\,,\label{La1}
\end{equation} 
and the dependence of $L_0$ is now hidden in the Lagrangian. Note that $L_0$ is the coordinate length of the fiducial cell in $t$-direction. One can in principle define the physical length of it as follows \cite{Bodendorfer:2019cyv}
\begin{equation}
\mathcal{L}_0\equiv\sqrt{a(r_0)}=L_0\sqrt{\bar{a}(r_0)}\,,
\end{equation}
where $r_0$ is a certain reference radius. The final physically relevant quantities are expected to be independent of these fiducial quantities $L_0$ and $\mathcal{L}_0$.

Using the Lagrangian \eqref{La1}, one can define the conjugate momenta $p_a$, $p_b$, and $p_n$. It can be seen that the conjugate momentum $p_n$ in the system \eqref{La1} is zero, indicating a primary constraint. This constraint is first class and it corresponds to a gauge degree of freedom. We can fix the gauge by choosing $n$ to be a constant.

In Ref.~\cite{Bodendorfer:2019cyv}, the authors introduced a new set of canonical conjugate variables
\begin{align}
v_1=\frac{2}{3}b^3\,,&\qquad P_1=\frac{a'}{\sqrt{n}b}=\left(\frac{p_b}{2b^2}-\frac{ap_a}{b^3}\right)\,,\nonumber\\
v_2=2ab^2\,,&\qquad P_2=\frac{b'}{\sqrt{n}b}=\frac{p_a}{2b^2}\,.\label{newcanonicalvar}
\end{align}
This set of variables relates to the original one ($a$, $b$, $p_a$, and $p_b$) via a canonical transformation. Using these new variables, the Hamiltonian constraint can be written in a much simpler form
\begin{equation}
H_{cl}=\sqrt{n}\mathcal{H}_{cl}\,,\qquad\mathcal{H}_{cl}=3v_1P_1P_2+v_2P_2^2-2\sim0\,.\label{classicalH}
\end{equation}
The $\sim$ denotes weak equality. 

After choosing the gauge $\sqrt{n}=\mathcal{L}_0$, the line element \eqref{SSSeq1} can be rewritten as
\begin{equation}
ds^2=-\frac{a(r)}{L_0^2}dt^2+\frac{\mathcal{L}_0^2}{a(r)}dr^2+b(r)^2d\Omega_2^2\,.\label{SSSeq2}
\end{equation}
The equations of motion can be obtained from the Hamiltonian. As a result, it can be proven that $b$ is proportional to $r$. In addition, one can recast $a(r)$ as a function of $b$ as follows
\begin{equation}
a(b)=\frac{\mathcal{L}_0^2}{\left(\frac{3D}{2}\right)^{\frac{2}{3}}}\left(1-\frac{F}{b}\right)\,,
\end{equation}
where $D$ is an integration constant and $F$ is a fiducial cell independent Dirac observable (constant only along the trajectories of the solution). After a constant $t$-rescaling and replacing $r$ with $b$, the classical Schwarzschild solution is recovered:
\begin{equation}
ds^2=-\left(1-\frac{2M}{b}\right)dt^2+\frac{1}{1-\frac{2M}{b}}db^2+b^2d\Omega_2^2\,,\label{SSSclassical}
\end{equation}
where the Dirac observable $F=2M$ is related to the mass of the black hole $M$.

\subsection{The effective model}
In Ref.~\cite{Bodendorfer:2019cyv}, the authors considered the polymerization of the classical model and solve the effective equations of motion to get the effective quantum Schwarzschild black hole. In this model, since the canonical variables directly relate to the metric functions of the whole spacetime, the effective equations of the whole spacetime, including the exterior of the event horizon, can be solved directly. Essentially, the effective Hamiltonian is derived by substituting the canonical momenta by the following sinusoidal functions:
\begin{equation}
P_1\rightarrow\frac{\sin\left(\lambda_1P_1\right)}{\lambda_1}\,,\qquad P_2\rightarrow\frac{\sin\left(\lambda_2P_2\right)}{\lambda_2}\,,\label{polymerizep1p2}
\end{equation}
where $\lambda_1$ and $\lambda_2$ are quantum parameters in this effective model. In the low-curvature regime where $\lambda_iP_i\ll1$, we have  $\sin(\lambda_iP_i)/\lambda_i\approx P_i$ and the effective model reduces to the classical model. However, the effective model would be significantly different compared with the classical model when the contributions from the quantum parameters become relevant. These significant quantum corrections are expected especially deep inside the black hole.

As just mentioned, the effective Hamiltonian is derived by substituting Eq.~\eqref{polymerizep1p2} into \eqref{classicalH}:
\begin{align}
H_{eff}&=\sqrt{n}\mathcal{H}_{eff}\,,\nonumber\\\mathcal{H}_{eff}&=3v_1\frac{\sin\left(\lambda_1P_1\right)}{\lambda_1}\frac{\sin\left(\lambda_2P_2\right)}{\lambda_2}+v_2\frac{\sin\left(\lambda_2P_2\right)^2}{\lambda_2^2}-2\nonumber\\ &\sim0\,.\label{effectiveH}
\end{align}
Solving the equations of motion from the effective Hamiltonian \eqref{effectiveH}, one can obtain $v_1(r)$, $v_2(r)$, $P_1(r)$, and $P_2(r)$. According to the relations in Eqs.~\eqref{newcanonicalvar}, the metric functions $b(r)$ and $a(r)$ in the line element \eqref{SSSeq2} can be solved as follows \cite{Bodendorfer:2019cyv}
\begin{widetext}
\begin{align}
b(r)&=\left(\frac{3v_1(r)}{2}\right)^{\frac{1}{3}}=\frac{\sqrt{n}}{\lambda_2}\left(3DC^2\lambda_1^2\right)^{\frac{1}{3}}\frac{\left(\frac{\lambda_2^6}{16C^2\lambda_1^2n^3}\left(\frac{\sqrt{n}r}{\lambda_2}+\sqrt{1+\frac{nr^2}{\lambda_2^2}}\right)^6+1\right)^{\frac{1}{3}}}{\left(\frac{\sqrt{n}r}{\lambda_2}+\sqrt{1+\frac{nr^2}{\lambda_2^2}}\right)}\,,\label{br}\\
a(r)&=\frac{v_2(r)}{2b(r)^2}=n\left(\frac{\lambda_2}{\sqrt{n}}\right)^4\left(1+\frac{nr^2}{\lambda_2^2}\right)\left(1-\frac{3CD}{2\lambda_2}\frac{1}{\sqrt{1+\frac{nr^2}{\lambda_2^2}}}\right)\frac{\left(\frac{1}{3DC^2\lambda_1^2}\right)^{\frac{2}{3}}\left(\frac{\sqrt{n}r}{\lambda_2}+\sqrt{1+\frac{nr^2}{\lambda_2^2}}\right)^2}{\left(\frac{\lambda_2^6}{16C^2\lambda_1^2n^3}\left(\frac{\sqrt{n}r}{\lambda_2}+\sqrt{1+\frac{nr^2}{\lambda_2^2}}\right)^6+1\right)^{\frac{2}{3}}}\,,\label{ar}
\end{align}
\end{widetext}
where $C$ and $D$ appear as two integration constants when solving $P_1(r)$ and $v_1(r)$, respectively. It should be noticed that when expressing the line element \eqref{SSSeq2}, the metric function $b(r)$ seems to be a better radial coordinate of the spacetime, rather than $r$. In fact, the metric functions $a(r)$ and $b(r)$ are well-defined for $r\in(-\infty,\infty)$. The existence of the negative branch of $r$ is absent in the classical model and it relates to an important feature of the effective spacetime: the existence of the white hole region. In Ref.~\cite{Bodendorfer:2019cyv}, the authors found that there are two fiducial cell independent Dirac observables in this effective spacetime, one corresponds to the black hole mass $M_{BH}$ and the other the white hole mass $M_{WH}$. By considering the approximated metric functions at large physical radius ($b(r)\rightarrow\infty$) as we will carry out later in section~\ref{subsec.approx}, the positive branch reduces to the Schwarzschild spacetime with a black hole mass $M_{BH}$. The negative branch, on the other hand, reduces to the Schwarzschild spacetime with a white hole mass $M_{WH}$ at its asymptotic region. These two masses are related to the two integration constants $C$ and $D$ as follows \cite{Bodendorfer:2019cyv}:
\begin{align}
C&=\frac{\lambda_2^3}{4\lambda_1\sqrt{n^3}}\left(\frac{M_{WH}}{M_{BH}}\right)^{\frac{3}{2}}\,,\nonumber\\ D&=\left(\frac{2\sqrt{n}}{\lambda_2}\right)^3\left[\frac{2}{3}\left(\frac{\lambda_1\lambda_2}{3}\right)^3M_{BH}^3\left(\frac{M_{BH}}{M_{WH}}\right)^{\frac{9}{2}}\right]^{\frac{1}{4}}\,.\label{CDmassrelation}
\end{align}
In fact the roles of two holes can be completely exchanged. As explained in Ref.~\cite{Bodendorfer:2019cyv}, an observer in the white hole exterior region would experience this region as the exterior of a Schwarzschild black hole spacetime.   

\subsection{The metric functions in terms of the radial coordinate $b$}
As we have mentioned, $r$ can be treated just as an auxiliary coordinate and it is $b(r)$ that has a meaning of physical radius. For practical purposes, it is more convenient to rewrite the metric functions in terms of $b$. This can be achieved by first inverting $b(r)$ to get $r=r(b)$, then using $a=a(r)$ to get $a(b)$.  

We first define the following variables:
\begin{equation}
y\equiv\frac{\sqrt{n}}{\lambda_2}r\,,\qquad X\equiv y+\sqrt{y^2+1}\,.\label{changev1}
\end{equation}
The metric functions $a(r)$ and $b(r)$ given in Eqs.~\eqref{br} and \eqref{ar} can be rewritten as functions of $X$ as follows
\begin{align}
b(X)&=A\frac{\left(BX^6+1\right)^{\frac{1}{3}}}{X}\,,\label{eq3}\\
a(X)&=\lambda_2^2\left(\frac{X^2+1}{2X}\right)^2\left(1-\frac{3CD}{2\lambda_2}\frac{2X}{X^2+1}\right)\frac{1}{b(X)^2}\,,\label{eq4}
\end{align}
where
\begin{equation}
A\equiv\frac{\sqrt{n}}{\lambda_2}\left(3DC^2\lambda_1^2\right)^{\frac{1}{3}}\,,\qquad
B\equiv\frac{\lambda_2^6}{16C^2\lambda_1^2n^3}\,.\label{eq6}
\end{equation}
Note that we have used the relation
\begin{equation}
y^2+1=\left(\frac{X^2+1}{2X}\right)^2\,,\nonumber
\end{equation}
which can be easily derived from the second equation in \eqref{changev1}.

Next, we invert Eq.~\eqref{eq3} to express $X$ as a function of $b$, i.e., $X=X(b)$. One can obtain
\begin{equation}
X(b)^3=\frac{b^3}{2A^3B}\pm\frac{1}{2}\sqrt{\frac{b^6}{A^6B^2}-\frac{4}{B}}\,.\label{eq7}
\end{equation}
In addition, from Eq.~\eqref{changev1} we get
\begin{align}
dr^2&=\frac{\lambda_2^2}{4n}\left(1+\frac{1}{X^2}\right)^2dX^2\nonumber\\
&=\frac{\lambda_2^2}{4n}\left(1+\frac{1}{X(b)^2}\right)^2\left(\frac{dX}{db}\right)^2db^2\,.\label{eq8}
\end{align}
Then, the metric line element \eqref{SSSeq2} can be written as
\begin{align}
ds^2=&-\frac{a(b)}{L_0^2}dt^2+\frac{\lambda_2^2}{4a(b)}\left(1+\frac{1}{X(b)^2}\right)^2\left(\frac{dX}{db}\right)^2db^2\nonumber\\&+b^2d\Omega^2_2\,.\label{SSSmetric3f}
\end{align}
Note that we have chosen the gauge $\sqrt{n}=\mathcal{L}_0$ as in the classical theory. As a result, the metric functions $g_{tt}$ and $g_{bb}$ in Eq.~\eqref{SSSmetric3f} can be written as functions of $b$ by using Eqs.~\eqref{eq4} and \eqref{eq7}.

To proceed, we insert the relations \eqref{CDmassrelation} into Eq.~\eqref{eq6} and get
\begin{align}
B&=\left(\frac{M_{BH}}{M_{WH}}\right)^3\,,\quad AB^{\frac{1}{3}}=\left[\frac{\lambda_1\lambda_2M_{BH}}{2}\left(\frac{M_{BH}}{M_{WH}}\right)^{\frac{3}{2}}\right]^{\frac{1}{4}},\nonumber\\
&CD=\frac{2}{\lambda_1}\left[\frac{2}{3}\left(\frac{\lambda_1\lambda_2}{3}\right)^3M_{BH}^3\left(\frac{M_{WH}}{M_{BH}}\right)^{\frac{3}{2}}\right]^{\frac{1}{4}}\,.\label{BCD}
\end{align}
It can also be seen that
\begin{equation}
\frac{3CDAB^{\frac{1}{3}}}{\lambda_2}=2M_{BH}\,,\qquad\frac{3CDA}{\lambda_2}=2M_{WH}\,.\label{holesmass}
\end{equation}
Finally, we choose a new time coordinate with a constant rescaling:
\begin{equation}
\tau\equiv\frac{\lambda_2}{2L_0AB^{\frac{1}{3}}}t\,,
\end{equation}
such that the metric \eqref{SSSmetric3f} can be written as
\begin{widetext}
\begin{equation}
ds^2=-\frac{4a(b)A^2B^{\frac{2}{3}}}{\lambda_2^2}d\tau^2+\frac{\lambda_2^2}{4a(b)}\left(1+\frac{1}{X(b)^2}\right)^2\left(\frac{dX}{db}\right)^2db^2+b^2d\Omega^2_2\,.\label{SSSmetric4f}
\end{equation}
\end{widetext}
This metric, Eq.~\eqref{SSSmetric4f}, in which $X(b)$ and $a(b)$ are defined through Eqs.~\eqref{eq7} and \eqref{eq4} respectively (see also Eq.~\eqref{changev1}), describes the effective metric of this polymer black hole. Later, we will use this metric \eqref{SSSmetric4f} to study the interior structure and QNMs of the BMM quantum black hole.

\subsection{The solutions at the asymptotic regions}\label{subsec.approx}

As can be seen in Eq.~\eqref{eq7}, the solution of the effective model contains a positive branch and a negative branch. In fact, the positive and negative branches can be regarded as the black hole region and the white hole region with different masses $M_{BH}$ and $M_{WH}$, respectively. This can be seen by considering the asymptotic expressions of the solution, i.e., at a large radius $b$.

We first consider the asymptotic region of the positive branch in which $y\rightarrow\infty$, $X\rightarrow\infty$, and $b\rightarrow\infty$. In this limit, according to Eqs.~\eqref{eq4} and \eqref{eq7}, $X(b)$ and $a(b)$ can be approximated as 
\begin{equation}
X(b)\approx\frac{b}{AB^{\frac{1}{3}}}\,,\qquad
a(b)\approx\frac{\lambda_2^2}{4A^2B^{\frac{2}{3}}}\left(1-\frac{3CD}{\lambda_2X(b)}\right)\,.
\end{equation}
Therefore, the metric \eqref{SSSmetric4f} can be approximated as
\begin{equation}
ds^2=-\left(1-\frac{2M_{BH}}{b}\right)d\tau^2+\frac{db^2}{1-\frac{2M_{BH}}{b}}+b^2d\Omega^2_2\,,
\end{equation}
where Eq.~\eqref{holesmass} has been used. As a consequence, the metric for the positive branch reduces to the Schwarzschild metric with a mass $M_{BH}$ at the asymptotic region.

On the other hand, the asymptotic region for the negative branch corresponds to $y\rightarrow-\infty$, $X\rightarrow0_+$, and $b\rightarrow\infty$. In this limit, the asymptotic expressions of $X(b)$ and $a(b)$ read
\begin{equation}
X(b)\approx\frac{A}{b}\,,\qquad
a(b)\approx\frac{\lambda_2^2}{4A^2}\left(1-\frac{3CDA}{\lambda_2b}\right)\,.
\end{equation}
Therefore, the metric \eqref{SSSmetric4f} can be written as
\begin{equation}
ds^2=-B^{\frac{2}{3}}\left(1-\frac{2M_{WH}}{b}\right)d\tau^2+\frac{db^2}{1-\frac{2M_{WH}}{b}}+b^2d\Omega^2_2\,,
\end{equation}
where Eq.~\eqref{holesmass} has been used again. After choosing a suitable constant time rescaling, the metric for the negative branch reduces to the Schwarzschild spacetime with a mass $M_{WH}$ at the asymptotic region.

It should be emphasized that the two branches of spacetime are smoothly connected when the square root in Eq.~\eqref{eq7} vanishes and the radius $b$ acquires its minimum value $b_m=\left(3\lambda_1CD/2\right)^{1/3}$. Replacing $CD$ with Eq.~\eqref{BCD}, we get
\begin{equation}
b_m=\left[2\left(\lambda_1\lambda_2\right)^3M_{BH}^3\left(\frac{M_{WH}}{M_{BH}}\right)^{\frac{3}{2}}\right]^{\frac{1}{12}}\,.
\end{equation}
Since $b$ gets its minimum value, the derivative $db/dX$ vanishes at $b=b_m$, indicating that the metric function $g_{bb}$ diverges at $b=b_m$. In addition, the metric function $g_{\tau\tau}$ acquires a non-zero finite value where the two branches of solution are connected. Actually, in this effective model, $b=b_m$ stands for a spacelike transition surface connecting a trapped and an anti-trapped regions, replacing the classical singularity \cite{Bodendorfer:2019cyv}.  

\subsection{Mass (de-)amplification relation}
The BMM quantum black hole is described by four parameters: two quantum parameters $\lambda_1$ and $\lambda_2$, and two distinct masses of the holes: $M_{BH}$ and $M_{WH}$. In Ref.~\cite{Bodendorfer:2019cyv} it has been shown that a certain mass (de-)amplification relation is required such that the condition that quantum effects become relevant only at a unique curvature scale which is independent of the mass of the black hole can be satisfied. Th mass (de-)amplification relation is \cite{Bodendorfer:2019cyv}: 
\begin{equation}
M_{WH}=M_{BH}\left(\frac{M_{BH}}{m}\right)^{\beta-1}\,,\qquad \beta=\frac{5}{3}\textrm{ or }\frac{3}{5}\,,\label{massamdeam}
\end{equation}
where the constant parameter $m$ has mass dimension and it is assumed to be smaller than the mass of the holes. If $\beta=5/3$, an observer traveling from the black hole to the white hole through the transition surface feels an amplification of the mass since $M_{WH}>M_{BH}$. On the other hand, if $\beta=3/5$ the observer would experience a mass de-amplification when crossing through the transition surface. It has been shown in Ref.~\cite{Bodendorfer:2019cyv} that this model is free from an indefinite mass amplification or de-amplification when crossing through multiple adjacent patches of Penrose diagram because the masses oscillate between $M_{BH}$ and $M_{WH}$ during the journey. The Penrose diagram of the full spacetime of the BMM quantum black hole was presented in Figure 8 of Ref.~\cite{Bodendorfer:2019cyv}. 

The reason and derivation of these two particular values of $\beta$ are clearly explained in Ref.~\cite{Bodendorfer:2019cyv}. This is essentially based on the requirement that the onset of the quantum effects should be at a unique curvature scale. To illustrate this, let us assume that the Kretschmann scalar $\mathcal{K}$ in this model can be roughly approximated as $\mathcal{K}\approx M^2/b^6$. This approximation is supposed to be valid at least at the regime where the quantum corrections are still moderate. Because the minimum radius at the interior region is given by $b_m=\left(3\lambda_1CD/2\right)^{1/3}$, provided the relation \eqref{massamdeam}, it can be shown that 
\begin{align}
b_m^3=\left[2\left(\lambda_1\lambda_2\right)^3\right]^{\frac{1}{4}}\times\begin{dcases} \frac{M_{BH}}{m^{\frac{1}{4}}} & \mbox{when }\beta=\frac{5}{3}  \\
\frac{M_{WH}}{m^{\frac{1}{4}}} & \mbox{when }\beta=\frac{3}{5}
\end{dcases}\,,\label{bm3cubic}
\end{align}
and the curvature scale $M^2/b^6$ close to the transition surface $b_m$ turns out to be independent of the mass on the de-amplified side, which is $M_{BH}$ ($M_{WH}$) when $\beta=5/3$ ($\beta=3/5$). On the amplified side, on the other hand, the curvature scale where quantum effects become important does depend on the mass ratio \cite{Bodendorfer:2019cyv}. But this mass dependence is irrelevant when the black hole is massive ($M\gg m$) since these quantum effects have significant contributions only deep inside the black hole. In the rest of this paper, we will rescale all quantities in the unit of $2M_{BH}$ such that $2M_{BH}=1$. Under this choice of rescaling, we will in addition assume $m=0.2$. 

\begin{figure*}[t]
\includegraphics[scale=0.55]{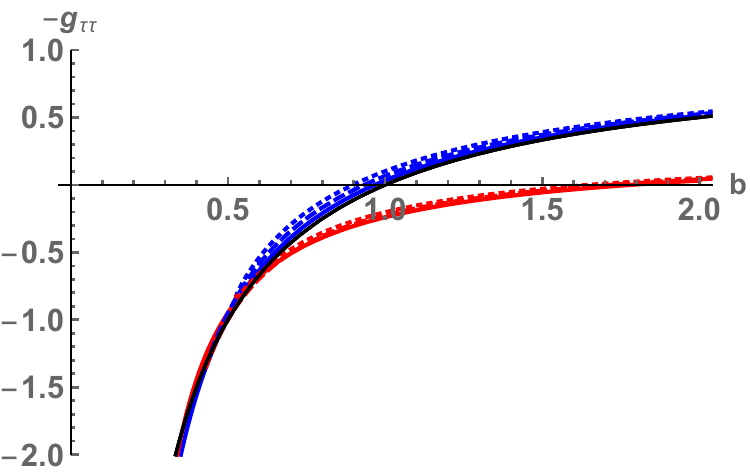}
\includegraphics[scale=0.55]{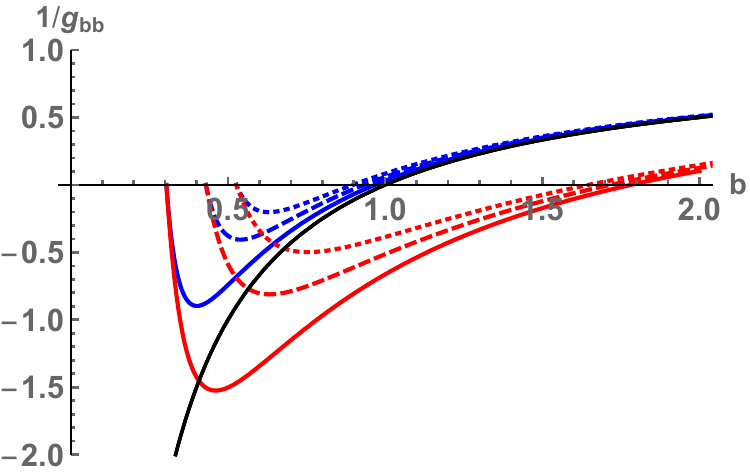}
\includegraphics[scale=0.55]{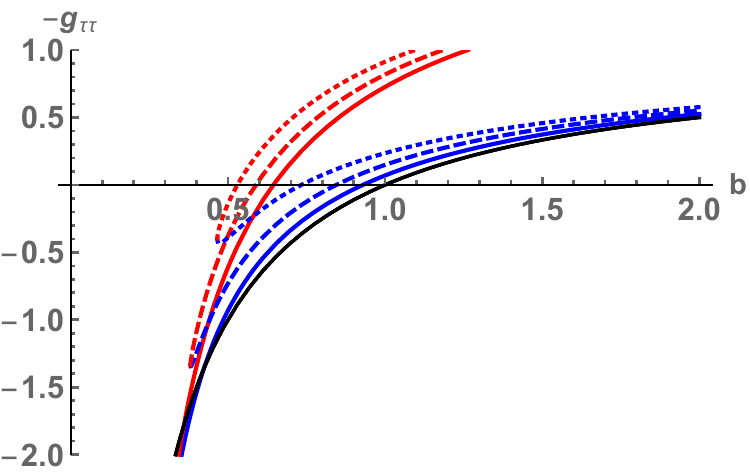}
\includegraphics[scale=0.55]{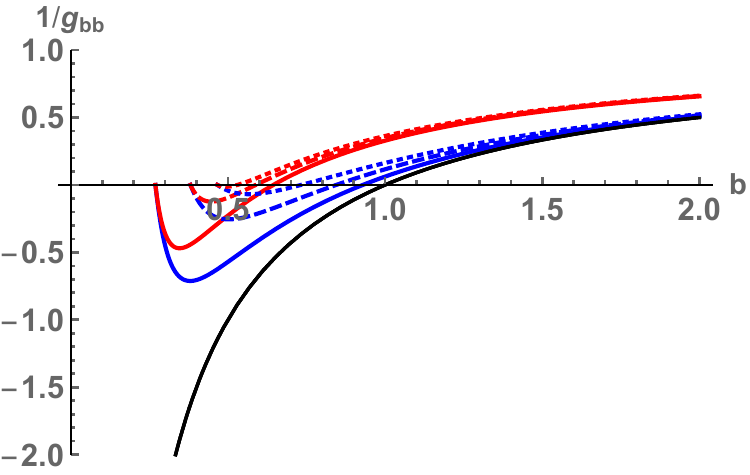}
\caption{\label{f1}The metric functions $-g_{\tau\tau}$ (left) and $1/g_{bb}$ (right) are shown in terms of $b$. The upper (lower) panel shows the results with $\beta=5/3$ ($\beta=3/5$). In each figure, the solid, dashed, and dotted curves correspond to $\lambda_1=\lambda_2=0.1$, $0.2$, and $0.3$, respectively. The Schwarzschild solution $1-1/b$ is presented by the black curves. One can see that the positive branch curve (blue) can be connected to the negative branch curve (red) at the transition surface $b_m$. Comparing the two branches, it can also be shown that the transition from the positive branch to the negative branch corresponds to a mass (de-)amplification when $\beta=5/3$ ($\beta=3/5$).}
\end{figure*}

In Figure~\ref{f1}, we have illustrated the metric functions of the BMM quantum black hole with respect to $b$. The left panel shows the metric function $-g_{\tau\tau}$ and the right panel shows $1/g_{bb}$. In the upper (lower) panel we have assumed $\beta=5/3$ ($\beta=3/5$). The blue curves are the positive branch solution, while the red curves represent the negative branch. We summarize some features which can be read off from Figure~\ref{f1}:
\begin{enumerate}[(i)]
\item The solutions reduce to the Schwarzschild metric (the black solid curves) at large radius or when the quantum parameters are small.
\item The size of the event horizon shrinks when the quantum parameters increase.
\item At the transition surface $b=b_m$, the metric function $g_{bb}$ diverges, as can be seen in the right panel.
\item The positive and the negative branches are connected at the transition surface. When $\beta=5/3$ ($\beta=3/5$), the negative branch (red curves) asymptotically reduces to a Schwarzschild metric with a mass $M_{WH}$ larger (smaller) than $M_{BH}$ after a constant time rescaling, corresponding to a mass (de-)amplification.
\end{enumerate}

\section{Interior structure}\label{sec.ts}
As we have mentioned, for the BMM quantum effective model, the black hole and white hole spacetimes are smoothly connected at a spacelike transition surface inside the event horizon. This transition surface takes place at a minimum radii $b=b_m$ where the metric function $g_{bb}$ diverges. The other metric function $g_{\tau\tau}$ is continuous and gets a finite value (see Figure~\ref{f1}). In this section, we will investigate features of the interior structure of this model, especially near the transition surface, in detail.

\subsection{Violation of energy conditions}\label{sec.VEC}
It can be imagined that the presence of transition surface replacing the singularity is purely due to the quantum effects in this model. If one uses an effective energy-momentum tensor to describe these quantum effects, this effective matter field should introduce some sorts of \textit{repulsive force} such that the singularity can be avoided. To reiterate, one does not assume any form of (phantom) matter or other such ingredients in order to ameliorate the singularity in this model; rather, one tries to write the quantum geometry corrections from the left hand side of Einstein's equation to the right hand side to reinterpret it as a sort of effective contribution to the stress energy tensor albeit the solution presented here is for LQG-modified \textit{vacuum} (spherically-symmetric) spacetime. Therefore, this `effective' matter field is expected to violate the energy condition. In this subsection, we shall confirm this hypothesis explicitly. 

As we have already discussed, we assume that the effective quantum black hole is described by the Einstein field equation with an effective energy-momentum tensor $T_{\mu\nu}$. The energy-momentum tensor can be written as that of an anisotropic fluid \cite{Herrera:1997plx}
\begin{equation}
T_{\mu\nu}=\left(\rho+p_2\right)u_\mu u_\nu+\left(p_1-p_2\right)x_\mu x_\nu+p_2 g_{\mu\nu}\,,\label{energymtensoranisotropic}
\end{equation}
where $\rho$ is the energy density measured by a comoving observer with the fluid, and $u^\mu$ and $x^\nu$ are the timelike four-velocity and the spacelike unit vector which is orthogonal to $u^\mu$ and the angular directions. In the expression \eqref{energymtensoranisotropic}, $p_1$ and $p_2$ are the radial pressure and the tangential pressure, respectively. Note that $u^\mu$ and $x^\mu$ satisfy
\begin{equation}
u_\mu u^\mu=-1\,,\qquad x_\mu x^\mu=1\,,\label{uuxx}
\end{equation}
where the indices are raised and lowered by the metric $g_{\mu\nu}$. Inside the horizon, The components of the energy-momentum tensor read
\begin{align}
T_{\tau\tau}&=g_{\tau\tau}p_1\,,\qquad T_\tau^\tau=p_1\,,\\
T_{bb}&=-g_{bb}\rho\,,\qquad T_b^b=-\rho\,,\\
&T_\theta^\theta=T_\phi^\phi=p_2\,.
\end{align}
The explicit expressions of $\rho$, $p_1$, and $p_2$ are functions of $b$ and they can be derived by calculating the corresponding Einstein tensor $G_{\mu\nu}(g)$ constructed from the metric Eq.~\eqref{SSSmetric4f}.

\begin{figure}
\includegraphics[scale=0.6]{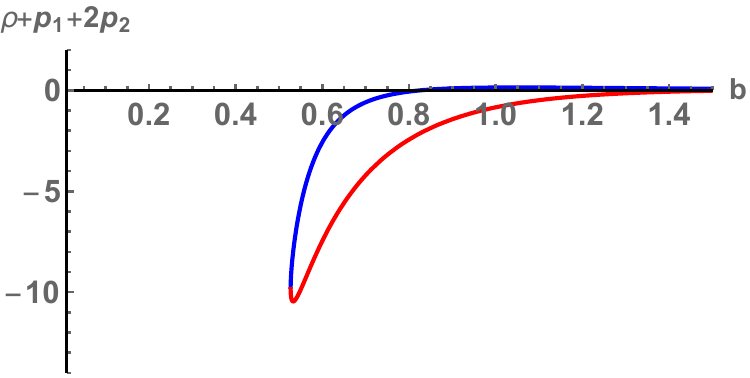}
\includegraphics[scale=0.6]{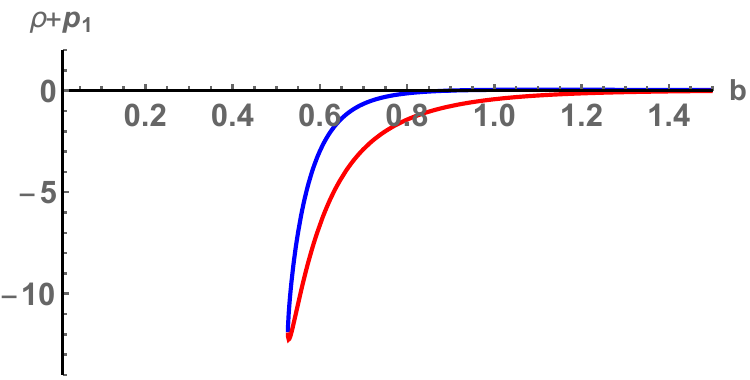}
\caption{\label{EC.fig}The violation of the energy conditions of the BMM quantum black hole is exhibited. Since $\rho+p_1+2p_2<0$ and $\rho+p_1<0$ at the transition surface, the strong energy condition and the null energy condition are violated. Here we assume $\beta=5/3$ and choose $\lambda_1=\lambda_2=0.3$. The blue curve corresponds to the positive branch of the solutions and the red curve corresponds to the negative branch.}
\end{figure}

For the energy-momentum tensor described by an anisotropic fluid, the strong energy condition is violated if and only if $\rho+p_1+2p_2<0$. On the other hand, the null energy condition is violated if and only if $\rho+p_i<0$ where $i=1,2$. According to Figure~\ref{EC.fig}, it can be seen that $\rho+p_1+2p_2$ and $\rho+p_1$ are negative near the transition surface in this model. Therefore, the strong and null energy conditions are violated for the BMM quantum black hole within the effective anisotropic fluid description.

\subsection{Junction conditions at the transition surface}
Since the BMM model contains a spacelike transition surface which connects two distinct spacetimes, it turns out to be crucial to investigate the junction condition at the transition surface more carefully. This can help us to check whether the transition of different spacetimes in this model is smooth or not. For instance, if the metric and its derivative acquire some discontinuities at the transition surface, one is obliged to include some non-trivial tension or pressure on this spacelike surface, which may lead to extra subtleties for this model.

In order to study the junction conditions on the transition surface in this model, we have to consider the junction conditions which are applicable i) in this effective quantum model, and, ii) for a spacelike hypersurface with the most general static and spherically symmetric metric. For the first point, since we are considering an effective black hole model based on LQG, it is fair and commonly acceptable to assume the validity of the Einstein's field equation (with an effective energy-momentum tensor) and therefore, the Israel junction conditions can be applied. Here we would like to mention that the junction condition and the equations of motion for general shells of arbitrary causal character (the character may even change) have been deduced and studied in arbitrary dimension \cite{Mars:1993mj,Mars:2007wy,Senovilla:2018hrw,Mars:2000gu,Mars:2007zt}, in which the Israel junction conditions are generalized. As for the second point, we will apply the Israel junction conditions to a spacelike hypersurface in the most general spherically symmetric metric. In fact, the junction condition for a spacelike hypersurface in a particular class of spherically symmetric metric has been deduced in Refs.~\cite{Balbinot:1990zz,Brahma:2018cgr}. In addition, the junction condition for a timelike hypersurface in the most general spherically symmetric metric has been investigated and applied in Refs.~\cite{BouhmadiLopez:2002mc,Garriga:1999bq,Garcia:2011aa,Bouhmadi-Lopez:2014gza,Kang:2019owv}. Here, we will consider the junction condition for a spacelike hypersurface embedded in the most general static and spherically symmetric spacetime, and apply it to the transition surface at the interior of the BMM quantum black hole.

We consider the following spacelike hypersurface 
\begin{equation}
ds_{t}^2=dz^2+b^2(z)d\Omega^2_2\,,\label{junctionh}
\end{equation}
which describes the junction between the positive and negative branches of the spacetime:
\begin{equation}
ds_{\pm}^2=-\left|g_{bb}^\pm\right| db^2+g_{\tau\tau}^\pm d\tau^2+b^2d\Omega^2_2\,.
\end{equation}
The functions $g_{bb}^\pm$ and $g_{\tau\tau}^\pm$ are the positive (negative) branches of the solution given by Eq.~\eqref{SSSmetric4f} evaluated on the two sides of the transition surface. Note that $g_{bb}<0$ and $g_{\tau\tau}>0$ inside the event horizon. 

The tangent vector on the junction surface can be constructed via the following vectors
\begin{align}
e_{z}^{\mu}&=\left(\dot{b},\quad\beta_\pm,\quad0,\quad0\right)\,,\nonumber\\
e_{\theta}^{\mu}&=\left(0,\quad 0,\quad 1,\quad 0\right)\,,\nonumber\\
e_{\phi}^{\mu}&=\left(0,\quad 0,\quad 0,\quad1\right)\,,\label{tangentve}
\end{align}
where $\beta_\pm\equiv\pm\sqrt{\left(1+\left|g_{bb}^\pm\right|\dot{b}^2\right)/g_{\tau\tau}^\pm}$ and the dots denote the derivative with respect to $z$. The vectors $e_a^\mu$ are constructed such that $h_{ab}=g_{\mu\nu}e_a^\mu e_b^\nu$ where $h_{ab}$ is the induced metric describing the junction surface \eqref{junctionh}. Using the vectors \eqref{tangentve}, the tangent vector $U^\alpha$ can be expressed using the $(b,\tau,\theta,\phi)$ coordinates as follows
\begin{equation}
U^\alpha=\left(\dot{b},\beta_\pm,0,0\right)\,.
\end{equation}
From the tangent vector $U^\alpha$, one can define a normal vector $n_\alpha$ such that $U^\alpha n_\alpha=0$ and $n^\alpha n_\alpha=-1$:
\begin{equation}
n_\alpha=\gamma_\pm\left(\beta_\pm,-\dot{b},0,0\right)\,,
\end{equation}
where $\gamma_\pm\equiv\sqrt{\left|g_{bb}^\pm\right|g_{\tau\tau}^\pm}$.

Furthermore, we calculate the extrinsic curvature $K_{ab}\equiv -n_{\mu;\nu}e_a^\mu e_b^\nu$ to get
\begin{equation}
K^a_b=\textrm{diag}\left[\frac{\dot{\left(g_{\tau\tau}^\pm\beta_\pm\right)}}{\gamma_\pm\dot{b}},\frac{\gamma_\pm\beta_\pm}{b\left|g_{bb}^\pm\right|},\frac{\gamma_\pm\beta_\pm}{b\left|g_{bb}^\pm\right|}\right]\,.
\end{equation}
Note that when $\gamma_\pm=1$, the result agrees with that in \cite{Balbinot:1990zz,Brahma:2018cgr}.

The shell stress energy tensor defined by a perfect fluid reads
\begin{equation}
S_b^a=\textrm{diag}\left[\sigma,P,P\right]\,,
\end{equation}
and it relates to the jump of the extrinsic curvature at the spacelike surface according to the Israel junction condition \cite{Israel:1966rt}:
\begin{equation}
8\pi S_{ab}=-\left(\left[K_{ab}\right]-h_{ab}\left[K\right]\right)\,,
\end{equation}
where $[K_{ab}]\equiv K_{ab}|^+-K_{ab}|^-$. Then we obtain
\begin{equation}
4\pi\sigma=\left[K_\theta^\theta\right]\,,\qquad 8\pi P=\left[K_\theta^\theta\right]+\left[K_z^z\right]\,.
\end{equation}

If we assume $\dot{b}=0$ at the junction surface, we obtain
\begin{align}
4\pi\sigma=&-\left(\frac{1}{b\sqrt{\left|g_{bb}\right|}}\right)\Bigg|^+-\left(\frac{1}{b\sqrt{\left|g_{bb}\right|}}\right)\Bigg|^-\,,\nonumber\\
4\pi P=&-\left(\frac{1}{2b\sqrt{\left|g_{bb}\right|}}+\frac{g_{\tau\tau,b}}{4g_{\tau\tau}\sqrt{\left|g_{bb}\right|}}\right)\Bigg|^+\nonumber\\&-\left(\frac{1}{2b\sqrt{\left|g_{bb}\right|}}+\frac{g_{\tau\tau,b}}{4g_{\tau\tau}\sqrt{\left|g_{bb}\right|}}\right)\Bigg|^-\,.\label{generaljc}
\end{align}
Eq.~\eqref{generaljc} is the junction condition for a spacelike hypersurface embedded in the most general static and spherically symmetric metric where $\gamma_\pm\ne1$. 

\begin{figure*}[tt]
\centering
\graphicspath{{fig/}}
\includegraphics[scale=0.46]{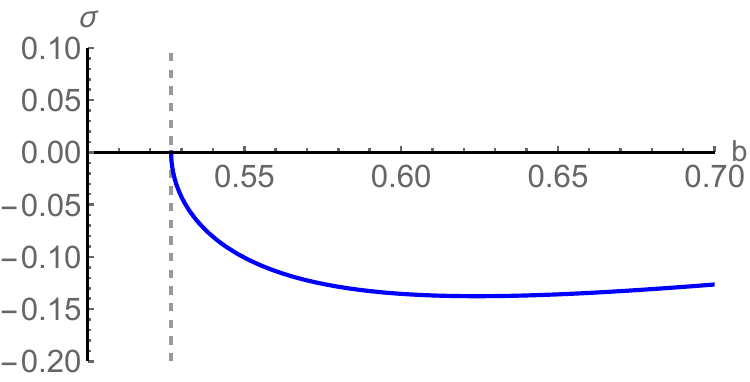}
\includegraphics[scale=0.46]{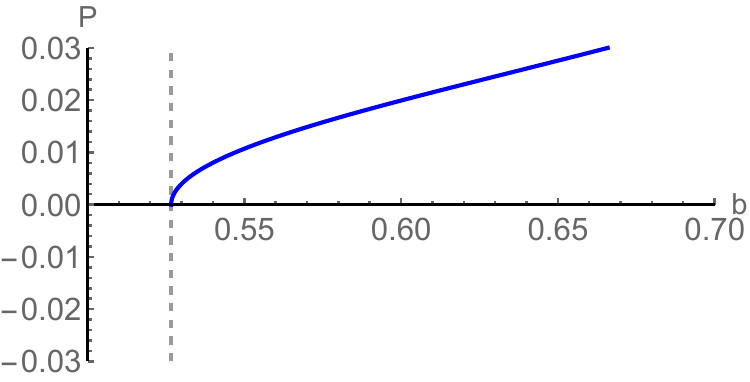}
\includegraphics[scale=0.46]{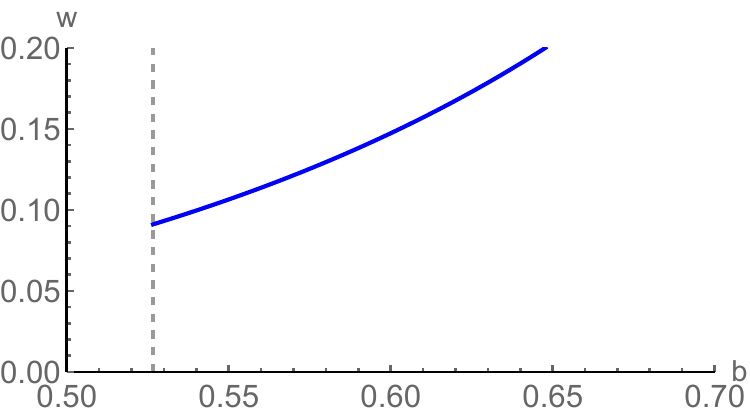}
\caption{The tension $\sigma$, pressure $P$, and the equation of state $w\equiv-P/\sigma$ on the junction surface. Here we assume $\beta=5/3$ and choose $\lambda_1=\lambda_2=0.3$. The dashed line corresponds to the bouncing point $b_m$.} 
\label{junction}
\end{figure*}

Since $\sqrt{g_{bb}}$ is proportional to $dX/db$ which diverges at the transition surface $b=b_m$, we find that $\sigma$ vanishes. As for the pressure $P$, we have found that the non-vanishing components of the extrinsic curvature on the two sides of the transition surface cancel with each other, therefore the pressure $P$ also vanishes. The equation of state $w\equiv -P/\sigma$, on the other hand, gets a positive finite value (see Figure~\ref{junction}). As a result, the vanishing of the tension $\sigma$ and the pressure $P$ clearly shows that the transition surface in the BMM black hole is indeed smooth and the two spacetimes are perfectly connected via this spacelike surface.

\section{Quasinormal modes}\label{sec.QNMbig}
In this section, we are going to study the perturbations of the BMM quantum black hole and their associated QNM frequencies. The QNMs associated with the gravitational perturbations of a black hole are tightly related to the ringing gravitational wave signals emitted from that black hole. The master equation describing the gravitational perturbations is derived by perturbing the spacetime metric, the energy-momentum tensor, and the gravitational equations. This is what we are going to carry out in section~\ref{sec.axialgqnm}. On the other hand, one can also consider the perturbation of some test fields, such as scalar fields or electromagnetic fields, in the given black hole spacetime. In this scenario, the notion of \textit{test fields} means that the back reaction of the test fields on the spacetime is assumed to be negligible. In this regard, the master equations are derived from the conservation equations of the associated test fields on the curved spacetime, such as the Klein-Gordon equation for a scalar field, and the Maxwell equations for electromagnetic fields. Later on, we will first consider the QNMs of the massless scalar field and the electromagnetic fields in sections~\ref{sec.masslessscalarfield} and \ref{sec.emqnm}, respectively. Then we investigate the axial gravitational perturbations and their QNMs in section~\ref{sec.axialgqnm}.

To derive the master equations of perturbations and study the QNMs of the BMM quantum black hole, in this work, we will consider the most general static and spherically symmetric spacetime where $g_{\tau\tau}g_{bb}\ne\textrm{constant}$. For the master equation of a massless scalar field, the derivation starting from the Klein-Gordon equation is rather straightforward. While for the electromagnetic and axial gravitational perturbations, we will resort to the tetrad formalism to obtain their master equations. 

Without loss of generality, the perturbed spacetime in our case can be described by a non-stationary and axisymmetric metric in which the symmetrical axis is turned such that no $\phi$ dependence appears in the metric functions. Considering only the axial perturbations, the perturbed metric $g_{\mu\nu}$ can be written as
\begin{align}
ds^2=&-\left|g_{\tau\tau}\right|d\tau^2+b^2\sin^2\theta\left(d\phi-\chi d\tau-q_2db-q_3d\theta\right)^2\nonumber\\
&+g_{bb}db^2+b^2d\theta^2\,,\label{generalmetricgaxial}
\end{align}
where the perturbations are encoded by the functions $\chi$, $q_2$, and $q_3$, and they are functions of time $\tau$, radial coordinate $b$, and polar angle $\theta$. The other metric functions $g_{\tau\tau}$ and $g_{bb}$ remain the zeroth order quantities and are functions of $b$ only.

Now, we use the tetrad formalism in which one defines a basis $e^\mu_{(a)}$ associated with the metric $g_{\mu\nu}$ \cite{Chandrabook}. The tetrad indices are enclosed in parentheses to distinguish them from the tensor indices. The tetrad basis should satisfy
\begin{align}
e_{\mu}^{(a)}e^{\mu}_{(b)}&=\delta^{(a)}_{(b)}\,,\quad e_{\mu}^{(a)}e^{\nu}_{(a)}=\delta^{\nu}_{\mu}\,,\nonumber\\
e_{\mu}^{(a)}&=g_{\mu\nu}\eta^{(a)(b)}e^{\nu}_{(b)}\,,\nonumber\\
g_{\mu\nu}&=\eta_{(a)(b)}e_{\mu}^{(a)}e_{\nu}^{(b)}\equiv e_{(a)\mu}e_{\nu}^{(a)}\,.
\end{align}
Conceptually, in the tetrad formalism we project the relevant quantities defined on the coordinate basis of $g_{\mu\nu}$ onto a chosen basis of $\eta_{(a)(b)}$ by constructing the tetrad basis correspondingly. In practice, $\eta_{(a)(b)}$ is usually assumed to be the Minkowskian matrix
\begin{equation}
\eta_{(a)(b)}=\eta^{(a)(b)}=\textrm{diag}\left(-1,1,1,1\right)\,.
\end{equation}
In this regard, any vector or tensor field can be projected onto the tetrad frame in which the field can be expressed through its tetrad components:
\begin{align}
A_{\mu}&=e_{\mu}^{(a)}A_{(a)}\,,\quad A_{(a)}=e_{(a)}^{\mu}A_{\mu}\,,\nonumber\\
B_{\mu\nu}&=e_{\mu}^{(a)}e_{\nu}^{(b)}B_{(a)(b)}\,,\quad B_{(a)(b)}=e_{(a)}^{\mu}e_{(b)}^{\nu}B_{\mu\nu}\,.
\end{align}
It should be emphasized that in the tetrad formalism, the covariant (partial) derivative in the original coordinate frame is replaced with the intrinsic (directional) derivative in the tetrad frame. For instance, the derivatives of an arbitrary rank two object $H_{\mu\nu}$ in the two frames can be related as follows \cite{Chandrabook}
\begin{align}
&\,H_{(a)(b)|(c)}\equiv e^{\lambda}_{(c)}H_{\mu\nu;\lambda}e_{(a)}^{\mu}e_{(b)}^{\nu}\nonumber\\
=&\,H_{(a)(b),(c)}\nonumber\\&-\eta^{(m)(n)}\left(\gamma_{(n)(a)(c)}H_{(m)(b)}+\gamma_{(n)(b)(c)}H_{(a)(m)}\right)\,,\label{2.7}
\end{align}
where a vertical rule and a comma denote the intrinsic and directional derivative with respect to the tetrad indices, respectively. A semicolon denotes a covariant derivative with respect to the tensor indices. Furthermore, the Ricci rotation coefficients are defined by
\begin{equation}
\gamma_{(c)(a)(b)}\equiv e_{(b)}^{\mu}e_{(a)\nu;\mu}e_{(c)}^{\nu}\,.
\end{equation}

\subsection{The massless scalar field perturbations}\label{sec.masslessscalarfield}
As the simplest example, we consider the QNMs of the massless scalar field around the BMM quantum black hole. The master equation describing the QNMs of the massless scalar field $\Phi$ is deduced from the Klein-Gordon equation in the curved spacetime:
\begin{equation}
\Box\Phi=\frac{1}{\sqrt{-g}}\partial_\mu\left(\sqrt{-g}g^{\mu\nu}\partial_\nu\Phi\right)=0\,.\label{KleinGordoneq}
\end{equation}
Since we have neglected the back reaction of the scalar field on the spacetime, we have treated the scalar field itself as a perturbation quantity. Therefore, we will only consider the zeroth order part of the metric \eqref{generalmetricgaxial}:
\begin{equation}
ds^2=-\left|g_{\tau\tau}\right|d\tau^2+g_{bb}db^2+b^2d\Omega^2_2\,.
\end{equation}
The scalar field $\Phi$ can be decomposed as follows:
\begin{equation}
\Phi(\tau,b,\theta,\phi)=\frac{1}{b}\sum_{l,m}\psi_l(\tau,b)Y_{lm}(\theta,\phi)\,,
\end{equation}
where $Y_{lm}$ are the spherical harmonics with $l$ and $m$ being the spherical harmonic indices. In this regard, the Klein-Gordon equation \eqref{KleinGordoneq} can be written as
\begin{equation}
\partial^2_{b_*}\psi_l+\omega^2\psi_l=V_s(b)\psi_l\,,\label{masslessscalarmaster}
\end{equation}
where $b_*$ refers to the tortoise radius defined by
\begin{equation}
\frac{db_*}{db}=\sqrt{\frac{g_{bb}}{\left|g_{\tau\tau}\right|}}\,.\label{tortoiser}
\end{equation}
Note that the Fourier decomposition ($\partial_\tau\rightarrow -i\omega$) has been used to obtain Eq.~\eqref{masslessscalarmaster}. The effective potential $V_s$ reads
\begin{equation}
V_s(b)=\left|g_{\tau\tau}\right|\left[\frac{l(l+1)}{b^2}+\frac{1}{b\sqrt{\left|g_{\tau\tau}\right|g_{bb}}}\left(\frac{d}{db}\sqrt{\frac{\left|g_{\tau\tau}\right|}{g_{bb}}}\right)\right]\,.
\end{equation}
The effective potential $V_s(b)$ with different parameters is shown in Figure~\ref{fsca}. The fundamental QNM frequencies are calculated using WKB method up to 6th order (see the appendix~\ref{WKBapp}) and the results are shown in Figure~\ref{QNMs}. We have also calculated the QNM frequencies using the AIM, and the results match those obtained through the WKB method. Note that we have presented the frequency ratio between the BMM quantum black hole and the Schwarzschild black hole, whose QNM frequency is denoted as $\omega_s$ \cite{Regge:1957td}.

\begin{figure}
\includegraphics[scale=0.55]{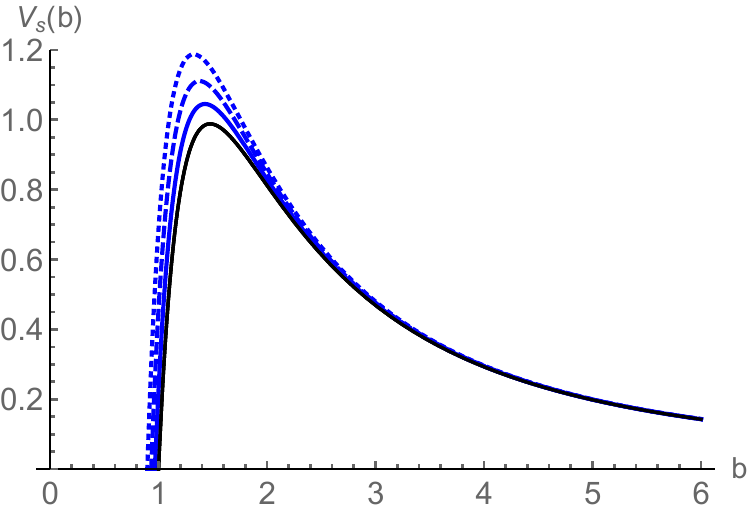}
\includegraphics[scale=0.55]{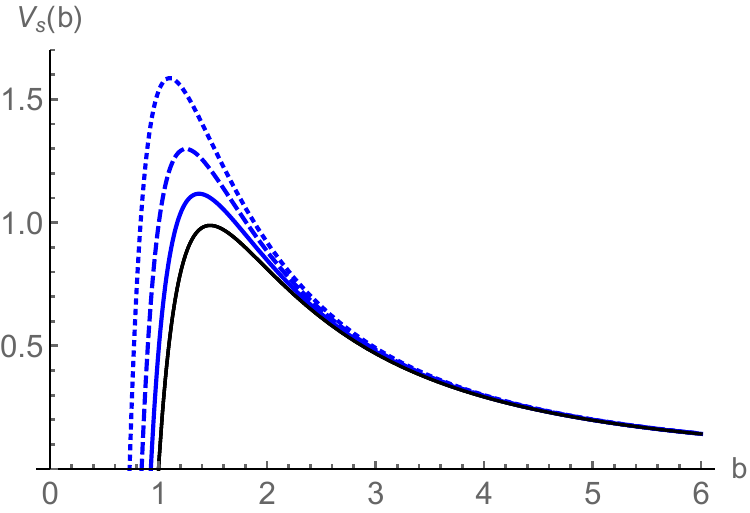}
\caption{\label{fsca}The effective potential $V_s(b)$ is shown for different values of parameters. The upper (lower) panel shows the results for $\beta=5/3$ ($\beta=3/5$). The solid, dashed, and dotted curves correspond to $\lambda_1=\lambda_2=0.1$, $0.2$, and $0.3$, respectively. The potential corresponding to the Schwarzschild solution is presented by the black curves. Here we assume the multipole number $l=2$.}
\end{figure}

\begin{figure*}[tt]
\centering
\graphicspath{{fig/}}
\includegraphics[scale=0.6]{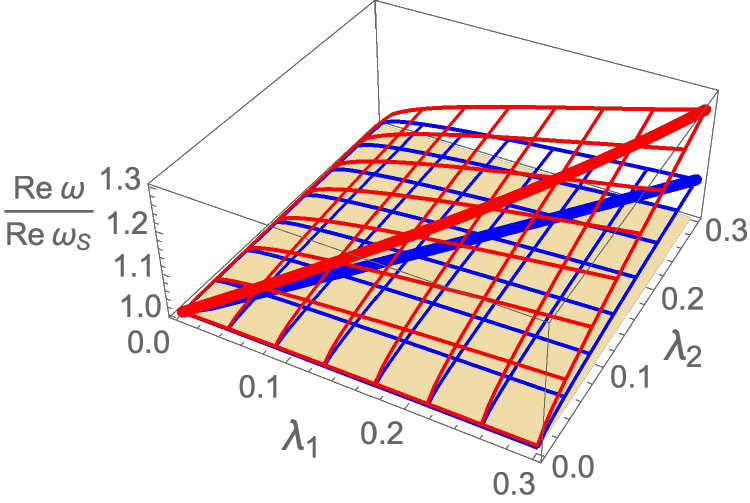}
\includegraphics[scale=0.5]{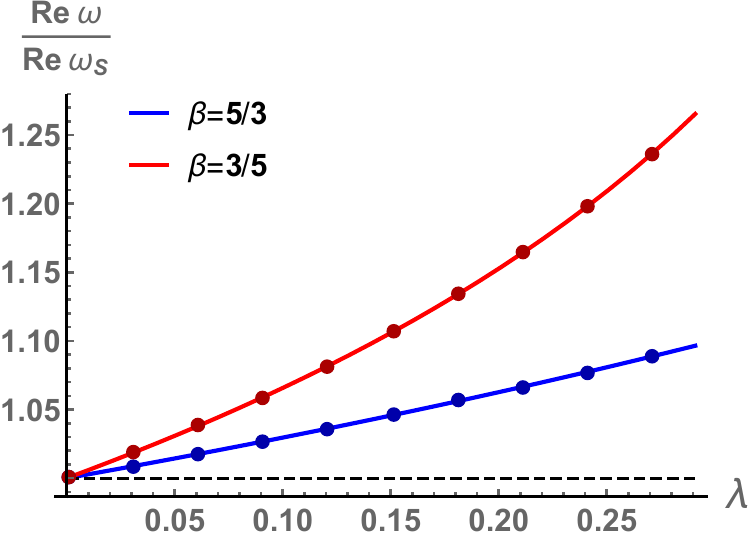}
\includegraphics[scale=0.6]{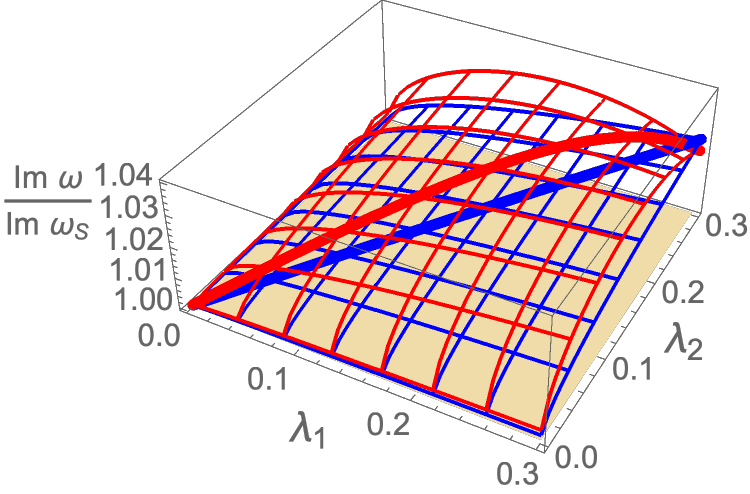}
\includegraphics[scale=0.5]{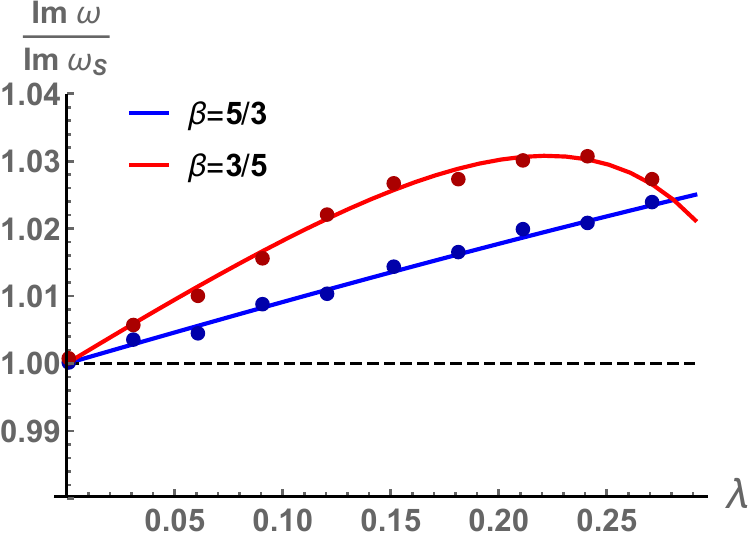}
\caption{\label{QNMs}The real part (upper) and the imaginary part (lower) of the fundamental QNMs for the massless scalar field are presented. Different colors represent different values of $\beta$. The left panel shows how the frequencies change with respect to the changes of $\lambda_1$ and $\lambda_2$. We highlight the curves of $\lambda_1=\lambda_2$ and show them in the right panel. In the right panel, the QNM frequencies evaluated using the AIM are shown as colored points, and they match well with the WKB results (solid curves). Here we assume the multipole number $l=2$.} 
\end{figure*}

\subsection{The electromagnetic perturbations}\label{sec.emqnm}
Next we consider the QNMs of the electromagnetic fields around the BMM quantum black hole. The master equation describing the electromagnetic QNMs is deduced from Maxwell equations. 

In the tetrad formalism \cite{Chandrabook}, the Bianchi identity of the field strength $F_{[(a)(b)|(c)]}=0$ gives
\begin{align}
&\left(b\sqrt{\left|g_{\tau\tau}\right|}F_{(\tau)(\phi)}\right)_{,b}+b\sqrt{g_{bb}}F_{(\phi)(b),\tau}=0\,,\label{bianchiEM1}\\
&b\sqrt{\left|g_{\tau\tau}\right|}\left(F_{(\tau)(\phi)}\sin\theta\right)_{,\theta}+b^2\sin\theta F_{(\phi)(\theta),\tau}=0\,.\label{bianchiEM2}
\end{align}
On the other hand, the conservation equation $\eta^{(b)(c)}(F_{(a)(b)})_{|(c)}=0$ gives
\begin{align}
\left(b\sqrt{\left|g_{\tau\tau}\right|}F_{(\phi)(b)}\right)_{,b}&+\sqrt{\left|g_{\tau\tau}\right|g_{bb}}F_{(\phi)(\theta),\theta}\nonumber\\&+b\sqrt{g_{bb}}F_{(\tau)(\phi),\tau}=0\,.\label{consEM}
\end{align}
For the sake of abbreviation, we then define the field perturbation
\begin{equation}
\mathcal{B}\equiv F_{(\tau)(\phi)}\sin\theta\,.
\end{equation}
After differentiating Eq.~\eqref{consEM} with respect to $\tau$ and using Eqs.~\eqref{bianchiEM1} and \eqref{bianchiEM2}, we have
\begin{align}
\left[\sqrt{\frac{\left|g_{\tau\tau}\right|}{g_{bb}}}\left(b\sqrt{\left|g_{\tau\tau}\right|}\mathcal{B}\right)_{,b}\right]_{,b}&+\frac{\left|g_{\tau\tau}\right|\sqrt{g_{bb}}}{b}\left(\frac{\mathcal{B}_{,\theta}}{\sin\theta}\right)_{,\theta}\sin\theta\nonumber\\&-b\sqrt{g_{bb}}\mathcal{B}_{,\tau\tau}=0\,.\label{EMmas1}
\end{align}
To proceed, we consider the Fourier decomposition and the following field decomposition \cite{Chandrabook}:
\begin{equation}
\mathcal{B}(b,\theta)=\mathcal{B}(b)Y_{,\theta}/\sin\theta\,,
\end{equation}
where $Y(\theta)$ is the Gegenbauer function satisfying \cite{Abramow}
\begin{equation}
\sin\theta\frac{d}{d\theta}\left(\frac{1}{\sin\theta}\frac{d}{d\theta}\frac{Y_{,\theta}}{\sin\theta}\right)=-l(l+1)\frac{Y_{,\theta}}{\sin\theta}\,.
\end{equation}
As a result, Eq.~\eqref{EMmas1} can be written as
\begin{align}
\left[\sqrt{\frac{\left|g_{\tau\tau}\right|}{g_{bb}}}\left(b\sqrt{\left|g_{\tau\tau}\right|}\mathcal{B}\right)_{,b}\right]_{,b}&+\omega^2b\sqrt{g_{bb}}\,\mathcal{B}\nonumber\\&-\frac{\left|g_{\tau\tau}\right|\sqrt{g_{bb}}}{b}l(l+1)\mathcal{B}=0\,.\label{EMmas2}
\end{align}

Finally, we redefine $\psi_{EM}\equiv b\sqrt{\left|g_{\tau\tau}\right|}\mathcal{B}$ and use the tortoise radius defined by \eqref{tortoiser}. The above equation \eqref{EMmas2} can be further written as
\begin{equation}
\partial_{b_*}^2\psi_{EM}+\omega^2\psi_{EM}=V_e(b)\psi_{EM}\,,\label{masterEMfieldequation}
\end{equation}
where the effective potential takes the following form \cite{Chandrabook}
\begin{equation}
V_e(b)=\left|g_{\tau\tau}\right|\frac{l(l+1)}{b^2}\,.
\end{equation}
The effective potential $V_e(b)$ with different parameters is presented in Figure~\ref{fe}. The fundamental QNM frequencies are calculated using the WKB method up to 6th order (see the appendix~\ref{WKBapp}) and the results are shown in Figure~\ref{QNMe}. Again, we have also calculated the QNM frequencies using AIM and the results match perfectly with those obtained through the WKB method. Note that we have presented the frequency ratio between the BMM quantum black hole and the Schwarzschild black hole, whose QNM frequency is denoted as $\omega_s$ \cite{Regge:1957td}.

\begin{figure}
\includegraphics[scale=0.55]{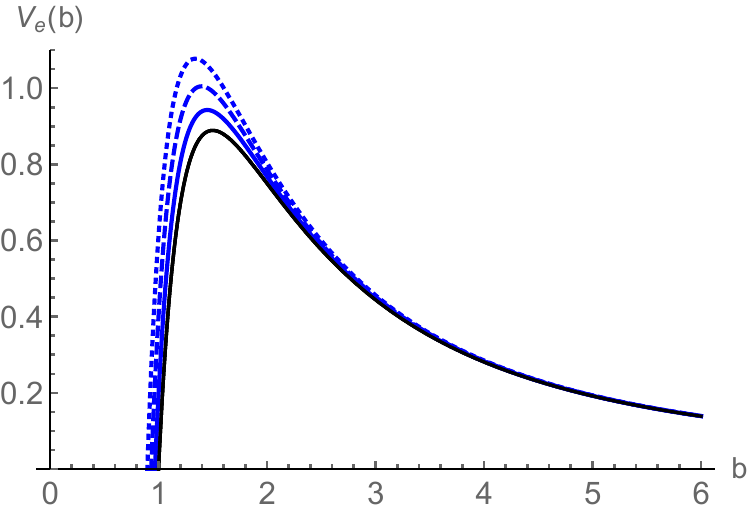}
\includegraphics[scale=0.55]{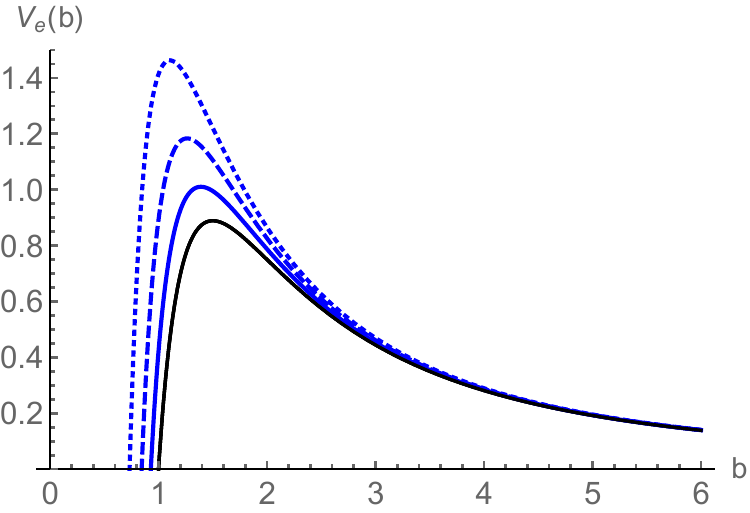}
\caption{\label{fe}The effective potential $V_e(b)$ is shown for different values of parameters. The upper (lower) panel represents the results for $\beta=5/3$ ($\beta=3/5$). The solid, dashed, and dotted curves correspond to $\lambda_1=\lambda_2=0.1$, $0.2$, and $0.3$, respectively. The effective potential for the Schwarzschild black hole is presented with the black curves. Here we assume the multipole number $l=2$.}
\end{figure}

\begin{figure*}[tt]
\centering
\graphicspath{{fig/}}
\includegraphics[scale=0.6]{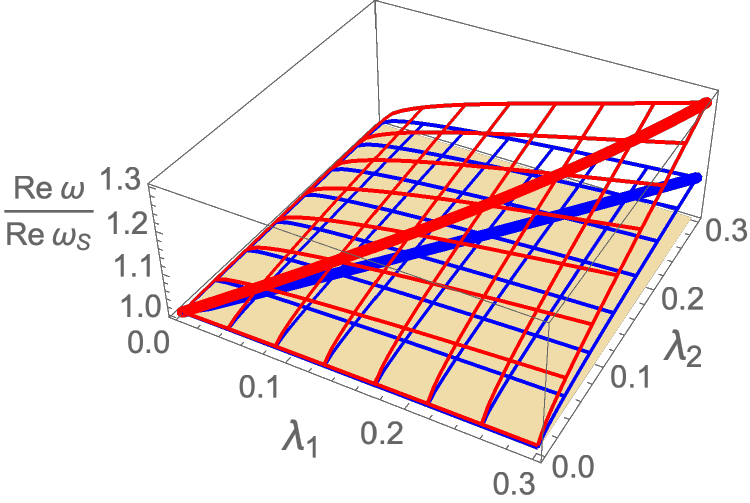}
\includegraphics[scale=0.5]{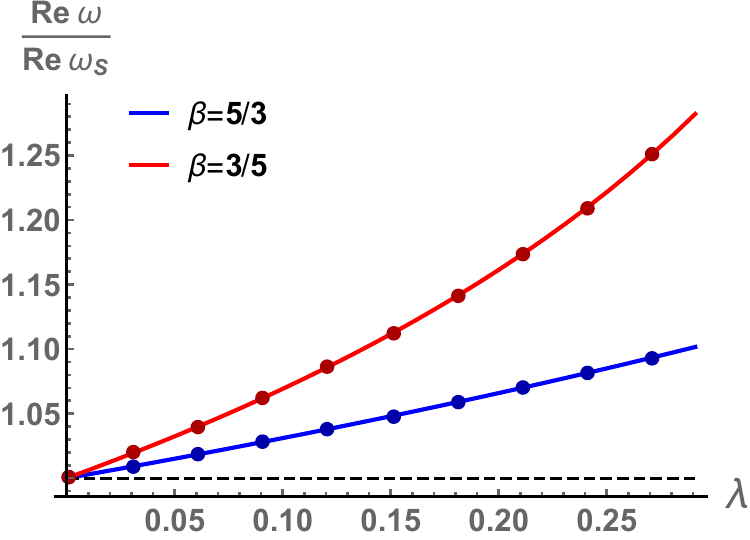}
\includegraphics[scale=0.6]{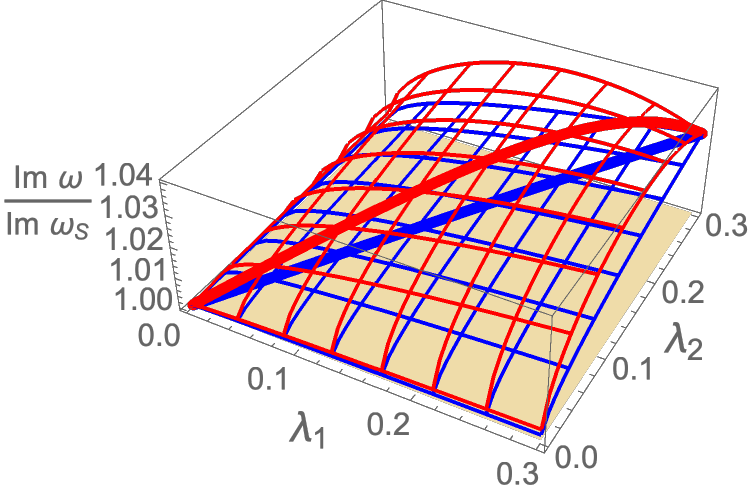}
\includegraphics[scale=0.5]{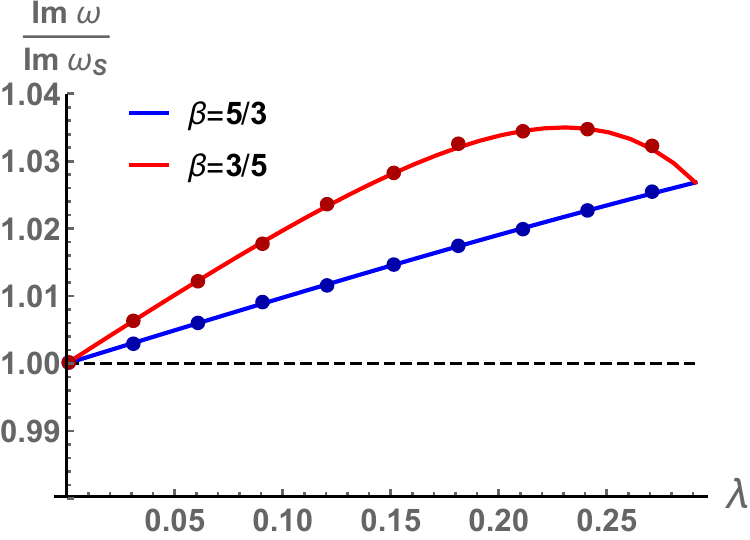}
\caption{\label{QNMe}The real part (upper) and the imaginary part (lower) of the fundamental QNMs for the electromagnetic perturbations are presented. Different colors represent different values of $\beta$. The left panel shows how the frequencies change with respect to the changes of $\lambda_1$ and $\lambda_2$. We highlight the curves of $\lambda_1=\lambda_2$ and show them in the right panel. In the right panel, the QNM frequencies evaluated using the AIM are shown as colored points, and it can be seen that they match well with the WKB results (solid curves). Here we assume the multipole number $l=2$.}
\end{figure*}

\subsection{The axial gravitational perturbations}\label{sec.axialgqnm}
In this subsection, we will consider the axial gravitational perturbations and their QNMs for the BMM quantum black hole. In fact, one has to perturb the corresponding gravitational equation and the energy-momentum tensor (if any) to derive the master equation. However, as we have mentioned, the BMM quantum black hole is just an effective model based on LQG in which the polymerization is performed at the level of Hamiltonian. Indeed, the \textit{modified Einstein's equation}, or the so-called \textit{effective equations} in LQG, can be derived from the effective Hamiltonian. However, instead of working directly with the effective equations,  we will apply a similar strategy to what we have used in section~\ref{sec.VEC}. We will assume that this effective quantum black hole is described by Einstein's gravity minimally coupled to an anisotropic fluid. Effectively, it is the anisotropic fluid that drives the quantum corrections. In this regard, we have to perturb the Einstein equation and the energy-momentum tensor of the anisotropic fluid. 

In the tetrad formalism, it has been proven in Ref.~\cite{Chen:2019iuo} that the axial components of the perturbed energy-momentum tensor defined by an anisotropic fluid are zero. Therefore, the master equation is given by the axial components of\footnote{It should be emphasized that the vanishing of the axial components of $R_{(a)(b)}$ is not equivalent to the vanishing of those of Ricci tensor $R_{\mu\nu}$ on the coordinate basis. In fact, the master equation of the axial perturbations derived from $R_{(a)(b)}=0$ is equivalent to that derived from the linearized Einstein equation coupled to the energy-momentum tensor of an anisotropic fluid, i.e., Eq.~\eqref{energymtensoranisotropic}.} $R_{(a)(b)}=0$. The $(\theta,\phi)$ and $(b,\phi)$ components of this equation read
\begin{align}
&\left[b^2\sqrt{\frac{\left|g_{\tau\tau}\right|}{g_{bb}}}\left(q_{2,\theta}-q_{3,b}\right)\right]_{,b}=b^2\sqrt{\frac{g_{bb}}{\left|g_{\tau\tau}\right|}}\left(\chi_{,\theta}-q_{3,\tau}\right)_{,\tau}\,,\label{grae1}\\
&\left[b^2\sqrt{\frac{\left|g_{\tau\tau}\right|}{g_{bb}}}\left(q_{3,b}-q_{2,\theta}\right)\sin^3\theta\right]_{,\theta}=\frac{b^4\sin^3\theta}{\sqrt{\left|g_{\tau\tau}\right|g_{bb}}}\left(\chi_{,b}-q_{2,\tau}\right)_{,\tau}\,,\label{grae2}
\end{align}
respectively. Then, we define
\begin{equation}
\mathcal{Q}\equiv b^2\sqrt{\frac{\left|g_{\tau\tau}\right|}{g_{bb}}}\left(q_{2,\theta}-q_{3,b}\right)\sin^3\theta\,.
\end{equation}
By differentiating Eqs,~\eqref{grae1} and \eqref{grae2} and eliminating $\chi$, we can get
\begin{align}
\left(\sqrt{\frac{\left|g_{\tau\tau}\right|}{g_{bb}}}\frac{\mathcal{Q}_{,b}}{b^2}\right)_{,b}&+\frac{\sqrt{\left|g_{\tau\tau}\right|g_{bb}}}{b^4}\left(\frac{\mathcal{Q}_{,\theta}}{\sin^3\theta}\right)_{,\theta}\sin^3\theta\nonumber\\&=\sqrt{\frac{g_{bb}}{\left|g_{\tau\tau}\right|}}\frac{\mathcal{Q}_{,\tau\tau}}{b^2}\,.\label{grae3}
\end{align}
We consider the Fourier decomposition and the ansatz
\begin{equation}
\mathcal{Q}(b,\theta)=\mathcal{Q}(b)Y(\theta)\,,
\end{equation}
where $Y(\theta)$ is the Gegenbauer function satisfying
\begin{equation}
\frac{d}{d\theta}\left(\frac{1}{\sin^3\theta}\frac{dY}{d\theta}\right)=-\left[l(l+1)-2\right]\frac{Y}{\sin^3\theta}\,.
\end{equation}
With these definitions, Eq.~\eqref{grae3} can be written as
\begin{align}
\left(\sqrt{\frac{\left|g_{\tau\tau}\right|}{g_{bb}}}\frac{\mathcal{Q}_{,b}}{b^2}\right)_{,b}&-\frac{\sqrt{\left|g_{\tau\tau}\right|g_{bb}}}{b^4}\left[l(l+1)-2\right]\mathcal{Q}\nonumber\\&=-\,\omega^2\sqrt{\frac{g_{bb}}{\left|g_{\tau\tau}\right|}}\frac{\mathcal{Q}}{b^2}\,.
\end{align}
Finally, we define $\psi_{G}\equiv\mathcal{Q}/b$ and consider the tortoise radius defined by Eq.~\eqref{tortoiser}. The master equation can be written as 
\begin{equation}
\partial_{b_*}^2\psi_{G}+\omega^2\psi_{G}=V_g(b)\psi_{G}\,,\label{mastergravitatioanlfieldeq}
\end{equation} 
where the effective potential reads
\begin{widetext}
\begin{equation}
V_g(b)=\left|g_{\tau\tau}\right|\left[\frac{l(l+1)}{b^2}+\frac{2\left(g_{bb}^{-1}-1\right)}{b^2}-\frac{1}{b\sqrt{\left|g_{\tau\tau}\right|g_{bb}}}\left(\frac{d}{db}\sqrt{\frac{\left|g_{\tau\tau}\right|}{g_{bb}}}\right)\right]\,.
\end{equation}
\end{widetext}
The effective potential $V_g(b)$ with different parameters is shown in Figure~\ref{fs}. The fundamental QNM frequencies are calculated using the WKB method up to 6th order (see the appendix~\ref{WKBapp}) and the results are shown in Figure~\ref{QNMg}. We have also calculated the QNM frequencies using AIM and the results match well with those obtained through the WKB method. Note that we have presented the frequency ratio between the BMM quantum black hole and the Schwarzschild black hole, whose QNM frequency is denoted as $\omega_s$ \cite{Regge:1957td}.

\begin{figure}[t]
\includegraphics[scale=0.55]{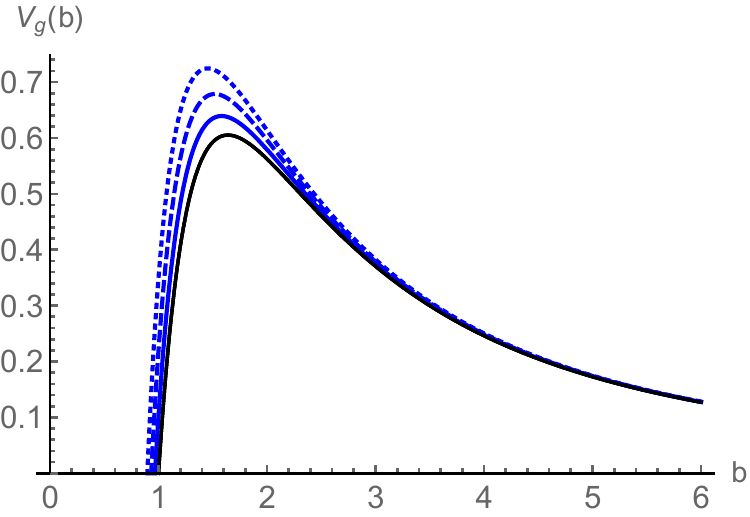}
\includegraphics[scale=0.55]{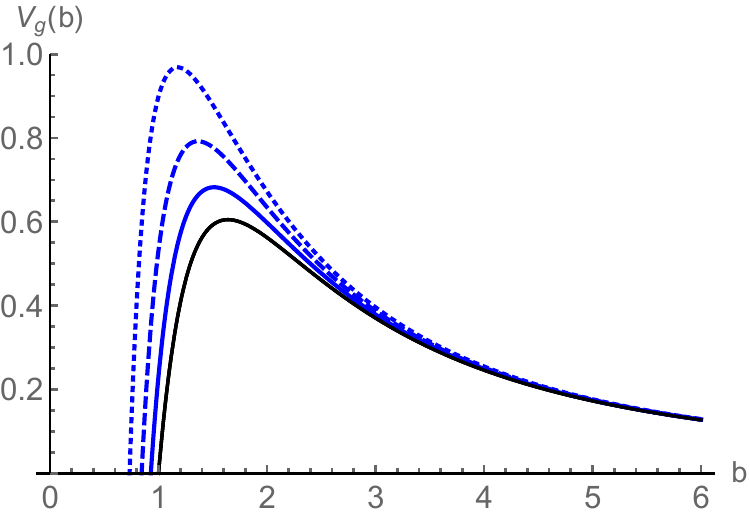}
\caption{\label{fs}The effective potential $V_g(b)$ is shown for different values of parameters. The upper (lower) panel shows the results for $\beta=5/3$ ($\beta=3/5$). The solid, dashed, and dotted curves correspond to $\lambda_1=\lambda_2=0.1$, $0.2$, and $0.3$, respectively. The effective potential for the Schwarzschild black hole is presented by the black curves. Here we assume the multipole number $l=2$.}
\end{figure}

\begin{figure*}[tt]
\centering
\graphicspath{{fig/}}
\includegraphics[scale=0.6]{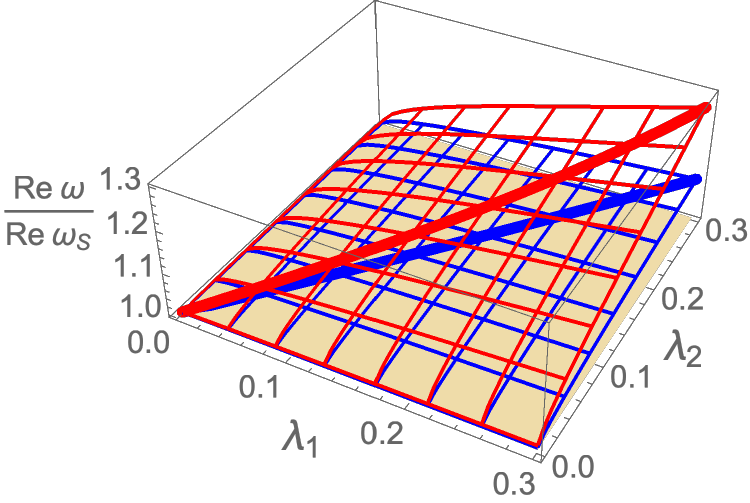}
\includegraphics[scale=0.5]{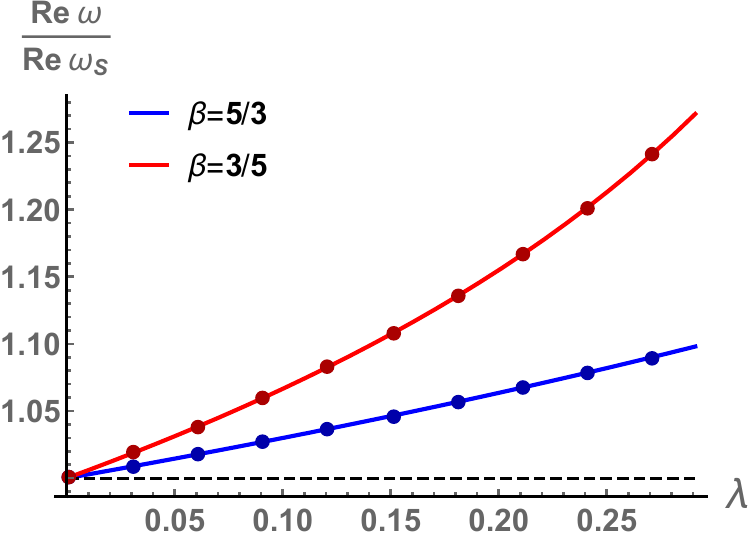}
\includegraphics[scale=0.6]{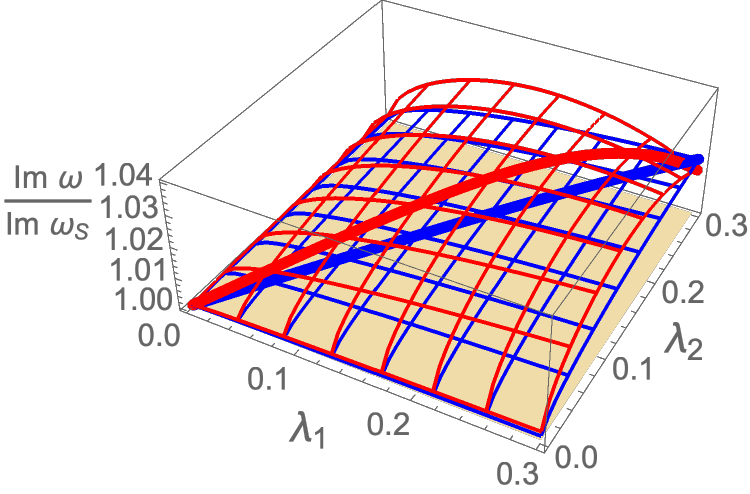}
\includegraphics[scale=0.5]{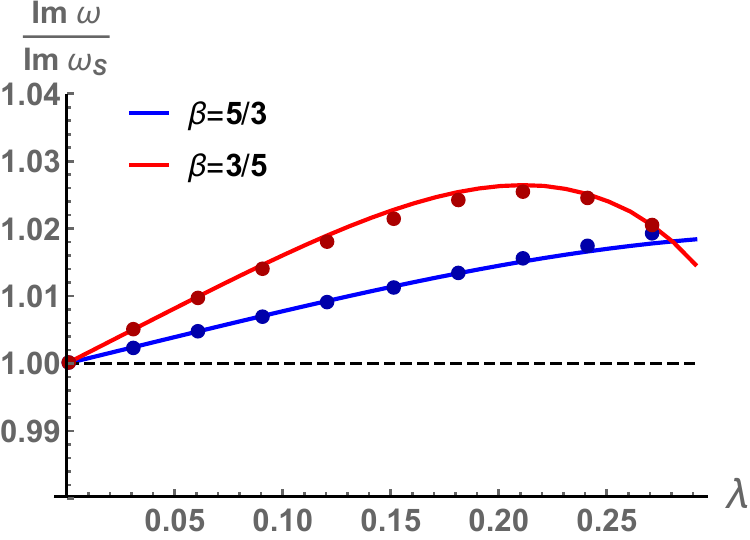}
\caption{\label{QNMg}The real part (upper) and the imaginary part (lower) of the fundamental QNMs for the axial gravitational perturbations are presented. Different colors represent different values of $\beta$. The left panel shows how the frequencies change with respect to the changes of $\lambda_1$ and $\lambda_2$. We highlight the curves of $\lambda_1=\lambda_2$ and show them in the right panel. In the right panel, the QNM frequencies evaluated using the AIM are shown as colored points, and they match well with the WKB results (solid curves). Here we assume the multipole number $l=2$.}
\end{figure*}

Let us briefly summarize what we have found in this section. First of all, we would like to emphasize that the BMM black holes with the parameter space chosen in this paper are linearly stable against the aforementioned perturbations. This is the consequence of the fact that the potentials (see Figures~\ref{fsca}, \ref{fe}, and \ref{fs}) are non-negative everywhere outside the horizon. Second, all these potentials share a similar qualitative shape, irrespective of the spin and the presence of the quantum parameters, although the height and the position of the peak would be different in different cases. Third, one can see that when changing the quantum parameters, the height of the potential increases and the radius of the event horizon shrinks. It is these changes that alter the QNM frequencies because the QNM frequencies are essentially determined by the potential. For example, if $\beta=3/5$, Figure~\ref{QNMe} shows that the real part of the QNM frequency is $15\%$ larger than the Schwarzschild counterpart when $\lambda_1=\lambda_2=0.2$ (note that we have rescaled $2M_{BH}=1$). The absolute value of the imaginary part, on the other hand, would increase by $3\%$. However, since the mass of a typical astrophysical black hole is much larger than the Planck mass, for this class of black holes, one should not expect significant deviations in observables between the BMM black hole and the Schwarzschild black hole.

\section{Conclusions}\label{conclu}

In this paper, we study the interior structure, the perturbations, and the QNMs of the BMM quantum black hole, which was proposed in Ref.~\cite{Bodendorfer:2019cyv}. Based on the framework of LQG, this effective black hole model is constructed via the polymerization of a new set of canonical phase space variables, in contrast to the standard SU($2$) connections and their conjugate momenta method commonly adopted in the LQG literature. The BMM quantum black hole is characterized by a spacelike transition surface inside the event horizon based on which the classical singularity is removed. In addition, the quantum effects rapidly die down when moving away from the transition surface and the solution recovers the Schwarzschild metric near and outside the event horizon.

First, we have proven that the energy conditions of the effective energy-momentum tensor are violated near the transition surface for the BMM quantum black hole. This is expected because such an effective matter field is supposed to provide some sort of \textit{repulsive force} inside the black hole in order to support the existence of the transition surface and to prevent the formation of singularity. This effective matter field is nothing but an effective description of the quantum corrections introduced in the LQG framework.

We then scrutinize the junction conditions on the transition surface. To do so, we applied the Israel junction condition to a spacelike hypersurface embedded in the most general static and spherically symmetric spacetime. We then investigated the junction condition on the transition surface of the BMM quantum black hole and found that the effective tension and pressure on the transition surface are zero, rendering a perfectly smooth transition between the black hole and the white hole regions. 

After studying the interior structure of the BMM quantum black hole, we investigated the black hole perturbations and the QNM frequencies in this model. For the massless scalar field perturbations and the electromagnetic perturbations, the master equations can be derived from the Klein-Gordon equation and the Maxwell equations, respectively. The master equation describing the axial gravitational perturbations, on the other hand, is obtained by linearizing the Einstein equation and the effective energy-momentum tensor of an anisotropic fluid. It can be proven that the axial components of the effective energy-momentum tensor in the tetrad frame vanish. Therefore, the master equation is derived from the axial components of the equation $R_{(a)(b)}=0$. We focused on the fundamental modes and calculated the QNM frequencies by using the WKB approach up to the 6th order as well as AIM. Our results show that each kind of perturbations shares the same qualitative tendency when changing the quantum parameters. That is, the real part of the QNM frequencies increases when the quantum parameters increase. On the other hand, the absolute value of the imaginary part of the QNM frequencies, which corresponds to the decay rate of the perturbations, increases as well when $\lambda_i$ increases slightly from zero. However, it would start to decrease when $\lambda_i$ acquires a larger value. Such deviations of QNM frequencies from those of the classical Schwarzschild black hole are due to the presence of quantum corrections and they are hopefully detectable with future advancement of gravitational wave observations. It should be emphasized that every physical quantity in this paper has been rescaled with the mass of the black hole ($2M_{BH}=1$). Therefore, the deviations of QNM frequencies from that based on GR would be more pronounced for smaller black holes. For astrophysical black holes, the deviations would be extremely small. For example, if we consider a solar mass black hole ($M_\odot\approx10^{38}M_P$), using Eqs.~\eqref{eq7} and \eqref{BCD}, we find that $\lambda_1\lambda_2$ has the dimension of $[M]^3$. Assuming that the quantum parameters satisfy $\lambda_1\lambda_2\approx M_P^3$, we find that $\lambda_1\lambda_2/M_{BH}^3=\lambda_1\lambda_2/M_\odot^3\approx 10^{-114}$, which is extremely small. Our estimation implies that the QNM spectra for LQG black holes can have important and detectable deviations only for Planckian black holes.

Although actual detection of these deviations from classical black holes would require a significant improvement in gravitational wave detection capabilities, our work establishes an interesting phenomenon: Quantum corrections which appear deep inside the interior of the black holes do `leak out' to the horizon, as evidenced by the QNM spectra of this model. This is in contrast to the effect induced by quantum corrections to the geometry of such effective black holes in LQG. As pointed out earlier, quantum modifications to the effective spacetime die out rapidly when moving away from the transition surface and geometric quantities such as curvature invariants recover their classical form near and outside the horizon. This implies that the geometric structure of such quantum-corrected black holes in LQG retains their classical properties near the event horizon\footnote{Indeed, this turns out to be a consistency requirement violated by some models of LQG black holes \cite{Brahma:2018cgr}.}. On the other hand, as we have shown, the QNMs of these models can still be sensitive to such quantum modifications, even if their magnitude would be necessarily suppressed for large mass black holes. In this sense, we have laid down the path, at least in principle, to falsify quantum corrections arising from LQG which are necessarily hidden deep inside the horizon. 

In conclusion, we have demonstrated that the BMM model not only is a consistent non-singular  black hole model in LQG and free from pathologies as pointed out in \cite{Brahma:2018cgr,Bodendorfer:2019xbp,Bouhmadi-Lopez:2019hpp,Bojowald:2019dry}, but also provides several signatures through their QNMs, which shall ultimately render this model distinguishable from GR. Furthermore, it would also be interesting to utilize the junction condition for spacelike hypersurfaces we have considered in this paper to investigate the properties of the transition surface of other quantum corrected black hole models. We leave these for our future works.
 
Note added: While we were preparing the final draft of this work, two new papers \cite{Bodendorfer:2019nvy,Bodendorfer:2019jay} appeared in which the authors of \cite{Bodendorfer:2019cyv} improved their model to make it free of its initial condition dependence, i.e., the requirement of certain values of $\beta$. We plan to study the QNMs and the interior structure of this model in the future.
 
\appendix 

\section{The WKB method for calculating QNM frequencies}\label{WKBapp} 
Essentially, the QNM frequencies are calculated by treating the master equations (Eqs.~\eqref{masslessscalarmaster}, \eqref{masterEMfieldequation}, and \eqref{mastergravitatioanlfieldeq} in this paper) as an eigenvalue problem with proper boundary conditions. In the literature, there have been various methods to calculate the QNMs, ranging from numerical approaches \cite{Leaver:1986gd,Jansen:2017oag} to semi-analytic methods (see Refs.~\cite{Nollert:1999ji,Berti:2009kk,Konoplya:2011qq,Berti:2015itd} and references therein). Among the plethora of technical methods, in this paper we have used a semi-analytical approach, which is constructed on the WKB approximation, to evaluate the QNM frequencies. The formulation of this method dates back to the seminal work \cite{Schutz:1985zz}. Furthermore, the 1st order WKB method has been extended to the 3rd and 6th order in Refs.~\cite{Iyer:1986np,Konoplya:2003ii}, respectively. Recently, a further extension of the WKB method up to the 13th order has been developed in Ref.~\cite{Matyjasek:2017psv,Matyjasek:2019eeu}. See Ref.~\cite{Konoplya:2019hlu} for a review on the developments of the WKB method to calculate QNM frequencies. 

By using the WKB method, the QNM frequencies can be directly evaluated with a simple formula as long as the effective potential in the master equation is known. It should be highlighted that the WKB method is accurate when the multipole number $l$ is larger than the overtone $\bold{n}$ \cite{Berti:2009kk}. Also, for astrophysical black holes, the fundamental modes have the longest decay time and would dominate the late time signal during the ringdown stage. These are the reasons why we have focused on the fundamental modes throughout this paper.

The idea of the WKB method relies on the boundary conditions that we need to impose when calculating QNM frequencies. At spatial infinity ($b_*\rightarrow\infty$), only outgoing waves moving away from the black hole exist. On the other hand, only ingoing waves moving toward the black hole can exist at the event horizon ($b_*\rightarrow-\infty$) because nothing can escape from the event horizon. To impose these boundary conditions, we treat the problem as a quantum scattering process without incident waves, while the reflected and the transmitted waves have comparable amounts of amplitudes. This can be achieved by assuming that the peak value of the effective potential satisfies $V(b)|_{\textrm{peak}}-\omega^2\gtrsim0$. There will be two classical turning points at the vicinity of the peak. At the regions far away from the turning points ($b_*\rightarrow\pm\infty$), the solution is obtained by using the WKB approximation up to a desired order, with the aforementioned boundary conditions. Near the peak, the differential equation is solved by expanding the potential into a Taylor series up to a corresponding order. After matching the solution near the peak with those derived from the WKB approximation simultaneously at the two classical turning points, the numerical values of the QNM frequencies $\omega$ can be deduced according to the matching conditions.

In the 6th order WKB method, the QNM frequencies can be evaluated with the following formula \cite{Schutz:1985zz,Iyer:1986np,Konoplya:2003ii}
\begin{equation}
\frac{i\left(\omega^2-V\right)}{\sqrt{-2\,\partial_{b_*}^2V}}\Bigg|_{\textrm{peak}}-\sum_{i=2}^6\Lambda_i=\bold{n}+\frac{1}{2}\,,\label{WKBformal}
\end{equation}
where $\Lambda_i$ are constant coefficients resulting from higher order WKB corrections. These coefficients contain the value and derivatives (up to the 12th order) of the potential at the peak. The explicit expressions of $\Lambda_i$ are given in Refs.~\cite{Iyer:1986np,Konoplya:2003ii} (see Eqs.~(1.5a) and (1.5b) in Ref.~\cite{Iyer:1986np}, and the appendix in Ref.~\cite{Konoplya:2003ii}).

\acknowledgments

SB thanks Norbert Bodendorfer for helpful comments. MBL is supported by the Basque Foundation of Science Ikerbasque. She also would like to acknowledge the partial support from the Basque government Grant No. IT956-16 (Spain) and from the project FIS2017-85076-P (MINECO/AEI/FEDER, UE). CYC and PC are supported by Ministry of Science and Technology (MOST), Taiwan, through No. 107-2119-M-002-005, Leung Center for Cosmology and Particle Astrophysics (LeCosPA) of National Taiwan University, and Taiwan National Center for Theoretical Sciences (NCTS). CYC is also supported by MOST, Taiwan through No. 108-2811-M-002-682. PC is in addition supported by US Department of Energy under Contract No. DE-AC03-76SF00515. The research of SB and DY is supported in part by the Ministry of Science, ICT \& Future Planning, Gyeongsangbuk-do and Pohang City and the National Research Foundation of Korea grant no. 2018R1D1A1B07049126. SB is also supported in part by funds from NSERC, from the Canada Research Chair program, by a McGill Space Institute fellowship and by a generous gift from John Greig.

\end{document}